\documentclass{mn2e}
\usepackage{graphicx,graphics}
\def\spose#1{\hbox to 0pt{#1\hss}}
\def\mdot {\spose{\raise 7.0pt\hbox{\hskip 5.0pt{\char '056}}}M}
\newcommand{\kms}{km\,s$^{-1}$}
\newcommand{\msol}{M$_\odot$}
\newcommand{\rsol}{R$_\odot$}

\newcommand{\ion}[2]{#1\,{\sc #2}}

\newcommand{\ph}{\phantom}
\begin{document}

\title
[Eclipsing binaries in the SMC]
{Ten eclipsing binaries in the Small Magellanic
Cloud:\\ fundamental parameters and SMC distance}
\author[Tim J. Harries, Ron W. Hilditch, and Ian D. Howarth]
{Tim J.~Harries$^1$, Ron W. Hilditch$^2$, and Ian D.~Howarth$^3$ \\
$^1$School of Physics, University of Exeter, Stocker Road, Exeter EX4
4QL, England. \\
$^2$School of Physics and Astronomy, University of St. Andrews, North
Haugh, St. Andrews, Fife KY16 9SS, Scotland. \\
$^3$Department of Physics and Astronomy, University College
London, Gower Street, London WC1E 6BT, England. \\
}
\date{Dates to be inserted}

\maketitle

\begin{abstract}
  We present the first results of an observational programme to
  measure the fundamental parameters of over 100 eclipsing binaries in
  the Small Magellanic Cloud (SMC). The spectroscopic data were
  obtained by using the 2dF multi-object spectrograph on the 3.9-m
  Anglo-Australian Telescope, and are used in conjunction with
  photometry from the OGLE database of SMC eclipsing binaries.  Ten
  systems are discussed in this first paper.  Three are detached
  early-B binaries, six are in a semi-detached configuration, and one
  is in a marginal contact state.  We conclude that the semi-detached
  systems are undergoing the slow mass-transfer phase of case-A binary
  evolution, in which the mass donor has reached its Roche lobe while
  still on the main sequence.  Each system provides a primary distance
  indicator.  By constructing a new calibration between spectral type
  and temperature for O and early B stars, we find a mean distance
  modulus to the SMC of $18.89 \pm 0.04$ (statistical) $\pm 0.10$
  (systematic). This value represents one of the most precise
  determinations to date of the distance to the SMC.

\end{abstract}


\section{Introduction}

The extragalactic distance scale is based on a small number of
calibrators, with Cepheids being the most widely adopted, primarily
due to the relative ease with which these objects may be identified
and measured in distant galaxies. However, other distance indicators,
such as red-clump and RR~Lyrae stars, retain a critically important
r\^ole in verifying the Cepheid distance scale.  Eclipsing,
double-lined spectroscopic binaries (eSB2s) not only provide a direct
route to fundamental determinations of stellar masses and radii, but
also afford true {\em primary} distance indicators if temperatures
(more precisely, surface fluxes) can be assigned to the components.
Our experience with studies of Galactic early-type binaries has shown
us that, with care, distance moduli accurate to $\pm0.1$~mag are
attainable (Hilditch, Harries \& Bell 1996; Harries, Hilditch \& Hill
1997; Harries, Hilditch \& Hill 1998), a precision comparable to that
obtained for individual Cepheid variables.

Distance determinations for eSB2s require densely sampled photometric
and spectroscopic data; such determinations have, therefore,
traditionally been observationally demanding, typically requiring
several nights of 4-m-class telescope time per object.  Two recent
developments have allowed significant advances in this area as far as
the Magellanic Clouds are concerned.  First, one of the valuable
legacies of the various microlensing surveys is a vast database of
newly discovered eclipsing binaries, with well-defined ephemerides and
well-sampled, high-quality, multicolour photometry; the OGLE survey
alone has yielded over 1400 new eclipsing binaries in the Small
Magellanic Cloud (SMC;  Udalski et al.\ 1998b).  Secondly, multi-object
fibre-fed spectrographs now allow data to be collected for many
objects simultaneously.  In this paper we report the first results of
a programme intended to determine fundamental parameters of $\sim$100
massive binaries in the SMC, by combining OGLE
photometry with spectroscopy obtained by using the 2dF multi-object
spectrograph on the Anglo-Australian Telescope (AAT).

The Magellanic Clouds constitute a crucial rung in the `distance
ladder', but recent determinations of their distance moduli (DMi)
have split between `short' and `long' scales (DMi $\sim$18.3 and
$\sim$18.5 for the LMC; e.g., Cole 1998).  The earliest attempt to
derive a DM for an individual eSB2 in the SMC was by Howarth (1982),
who found a DM of 18.3 for the X-ray binary Sk~160, located in the
wing of the SMC ($\sim$0.3 mag closer than the main body of the SMC
according to Caldwell \& Coulson 1986).  Bell et al.\ (1991) presented
the first eSB2-based DM for the SMC itself, using HV~2226 to estimate
a DM of $18.6\pm0.3$.  Guinan and co-workers are currently undertaking
a programme to study in detail individual B-star binaries in the LMC,
using HST spectra to determine temperatures and reddenings
simultaneously. To date, this work has led to distances to two
objects, HV~2274 (Guinan et al.\ 1998; Ribas et al.\ 2000) and HV~982
(Fitzpatrick et al.\ 2002). The DM found for HV~2274 ($18.35 \pm
0.07$) appears to support the short distance scale to the Clouds, but
the DM of HV~982 found by Fitzpatrick et al.\ (2002; 18.50) is
consistent with the long distance scale.  Re-analysis of the HV~2274
data by other authors has also resulted in larger DM estimates ($18.40
\pm 0.07$, Nelson et al.\ 2000; $18.46 \pm 0.06$, Groenewegen \&
Salaris 2001).

The aims of our programme are to provide a reliable eSB2-based
distance estimate for the SMC, and to determine masses, radii and
temperatures for $\sim$200 stars. These data will be used to test
low-metallicity stellar-evolution models for both single stars (the
detached systems) and interacting binary systems (the semi-detached
and contact objects).  The present paper is a first step in this
direction.  We summarize the selection criteria for the sample, and
the spectroscopic observations, in Section~\ref{obs_txt}. We then
describe the basic steps in the analysis, and tabulate our results for
the first ten programme objects. Finally, we report our initial
distance estimate, and compare our result with work based on red-clump
stars and the Cepheid period--luminosity relation.

\section{Observations}
\label{obs_txt}

We base our sample on the OGLE database of SMC eclipsing binaries
(Udalski et al.\ 1998).  The SMC stellar population is of intrinsic
interest as the most easily observed sample of low-metallicity
objects, and the quality of the OGLE photometry and astrometry is well
suited to our project.  To define the target sample, two filters were
initially applied to the OGLE eclipsing-binary database: $B <16$, to
allow sufficient signal-to-noise on our faintest targets ($\mbox{S:N}
\ga 25$ per wavelength sample); and orbital period $P_{\rm orb} < 5$d,
to ensure adequate phase coverage for most targets in our first
5-night allocation of observing time.

Our spectroscopy was obtained with the 2dF instrument, which is
capable of securing up to 400 spectra simultaneously, using 200 fibre
feeds to each of two spectrographs (Lewis et al.\ 2002).  It has a
2$^\circ$ field of view, and two substantially overlapping fields were
therefore required to ensure complete coverage of the 2.5 square
degree OGLE region.  The adopted field centres were 00$^{\rm
h}$\,47$^{\rm m}$\,00$^{\rm s}$ $-73^\circ$\,10$'$00$''$ (J2000,
herein Field~1), and 01$^{\rm h}$\,00$^{\rm m}$\,00$^{\rm s}$\
$-72^\circ$\,45$'$\,00$''$ (Field~2).  Astrometry accurate to better
than 0.5$''$ is essential for 2dF observations, since the fibre
diameter is only 2$''$. We cross-correlated the OGLE catalogue against
SuperCosmos source lists and found good agreement in relative
position, with a sub-arcsecond offset in right ascension and
declination. We therefore used the OGLE positions to assign the 2dF
fibre positions.   Fibre-allocation constraints (such as fibre crossings
and angular proximity of targets) meant that a few stars in the
filtered catalogue were unobservable.  The final target list consists
of 124 binaries, of which 69 were in the overlap region of the two
fields.

Digitized Sky Survey (DSS) images were examined to identify sky
regions in these densely populated fields, and the sky positions
manually allocated.  Bright ($14 < V < 15$) fiducial, or guide, stars
were selected from the full OGLE catalogue, and DSS images of the
objects were again inspected to ensure that there were no blended or
other close companions.

\begin{figure*}
\includegraphics{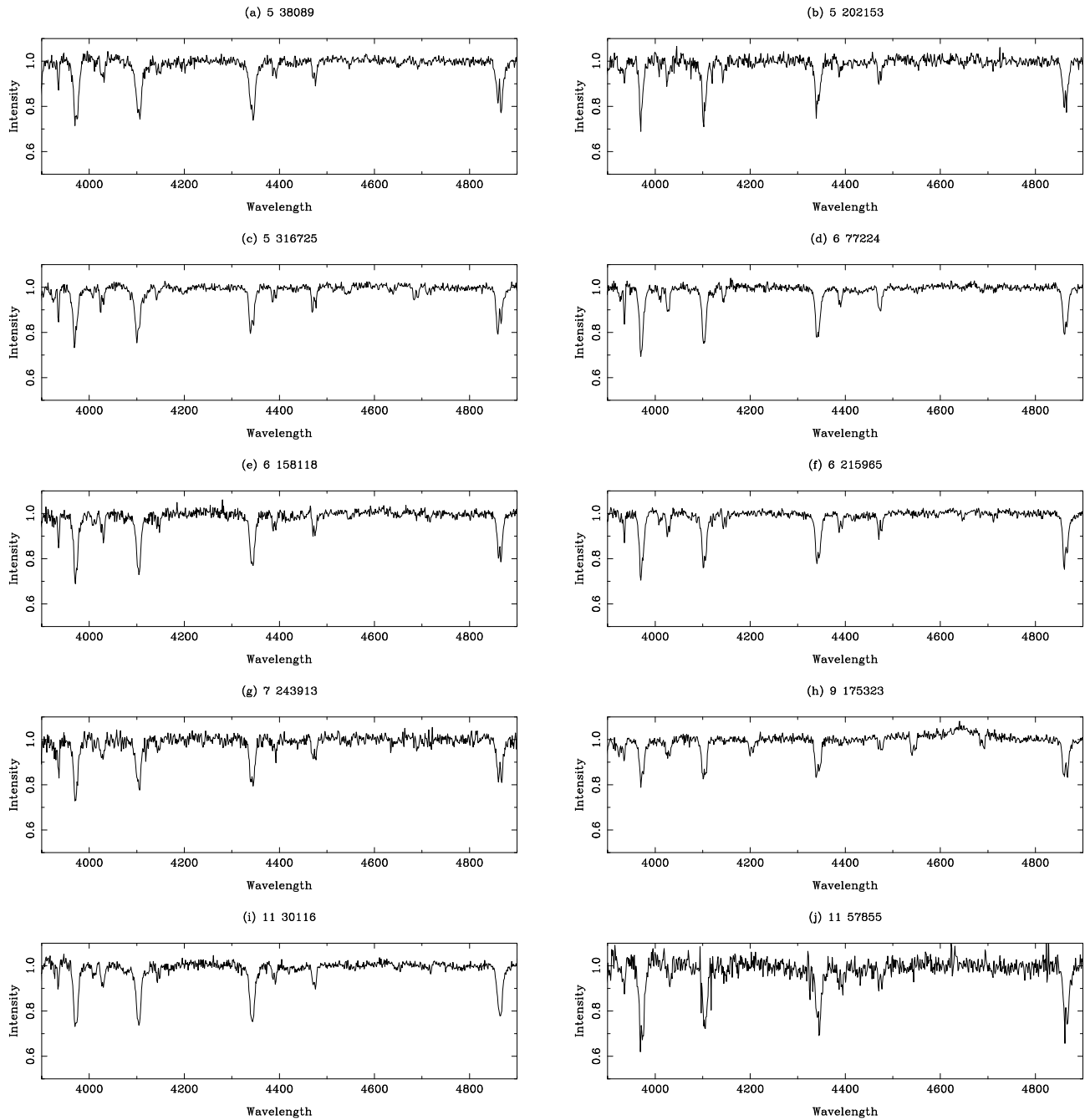}
\caption{Illustrative spectra of the targets discussed in this
  paper, taken close to maximum postive velocity of the primary
  (spectroscopic phase zero, photometric phase 0.75). The strong
  absorption lines visible include H$\delta$, H$\gamma$, H$\beta$,
  \ion{He}{i}~$\lambda 4471$.  Note that in several cases (e.g.,
  9~175323) the absorption lines of both components are clearly
  visible.}
\label{raw_spectra_fig}
\end{figure*}

The data were obtained in service mode by T.~Bridges during the nights
of 2001 Sept.~6--11 (JDs 2,452,158--63), with one test exposure of
Field~1 taken on Sept.~5; the night of Sept.~7 was lost to cloud.
Integration times of 1800s were used for most target exposures, in
seeing of typically 1.5--2 arcseconds. As each field contained fewer
than 200 targets, only the better-performing of the 2dF spectrographs
was employed. A 1200B grating was used, providing a mean reciprocal
dispersion of 1.1\AA\,px$^{-1}$, and a resolution of 2 \AA\ (135
\kms). The wavelength range of the data is 3855\AA\ to 4910\AA, a
region which contains spectral lines suitable for both radial-velocity
measurements and spectral typing in the classification regime of
interest.

Arc and flatfield frames were obtained after each target exposure,
along with offset sky frames (which may be used to calibrate fibre
throughput), and twilight-sky integrations.  The {\sc 2dFdr} package
was used to reduce the data, with default settings (Lewis et al.\
2002). Twilight-sky observations were used to calibrate the system
throughputs when possible. The final dataset comprises 32 exposures (16
per field), so each system has at least 16 orbital-phase points, with
rather more than half the targets having 32 phase points.

For the present study we selected ten binaries on the basis of
spectral signal-to-noise and coverage of quadrature phases (the times
of maximum radial-velocity separation, best constraining the orbital
velocity amplitudes). The selected objects are listed in
Table~\ref{basic_param_tab}, and examples of typical spectra are shown
in Figure~\ref{raw_spectra_fig}.

\begin{figure*}
\includegraphics{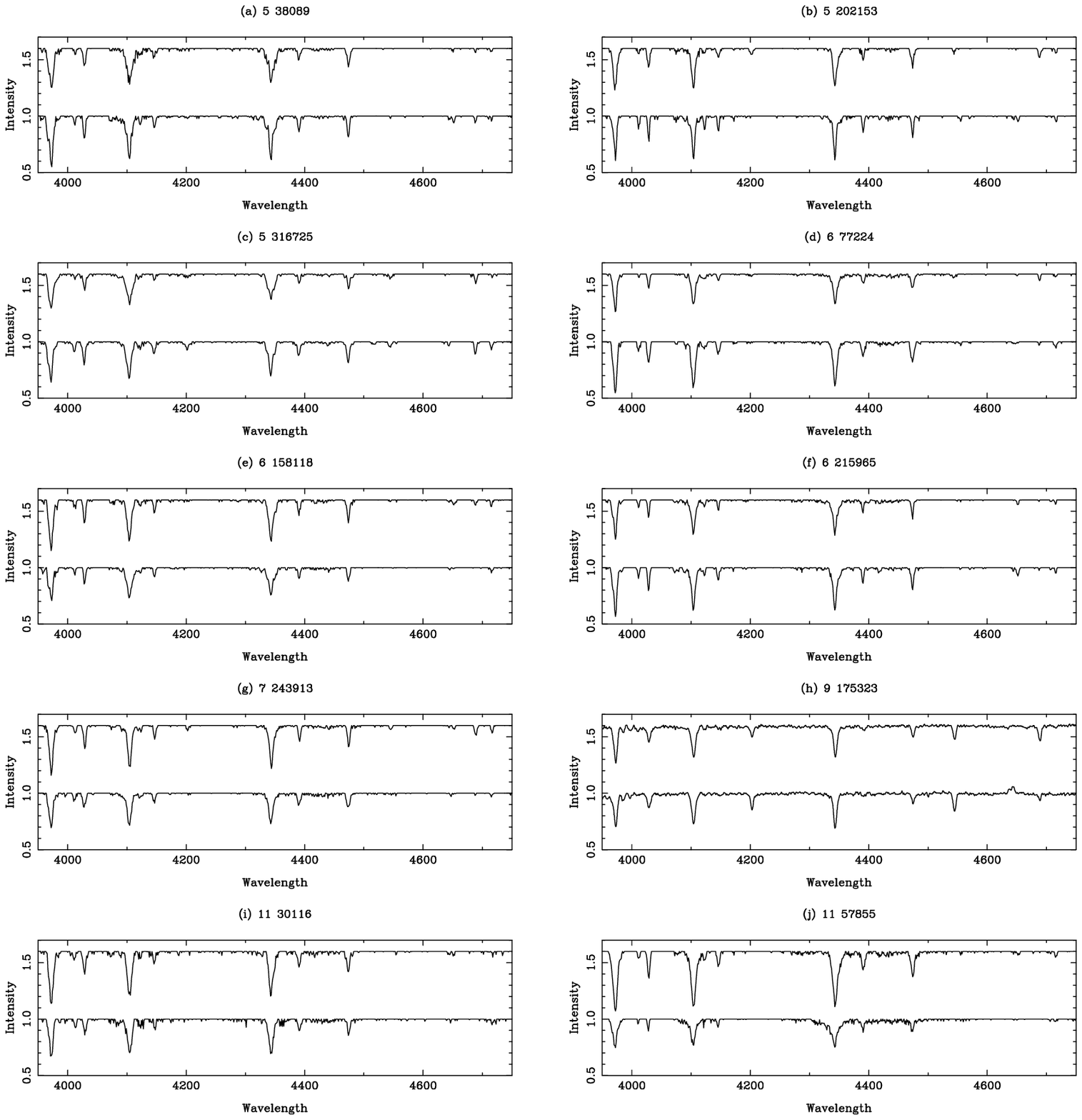}
\caption{Component spectra for the subsample. The normalized spectra
are plotted as relative intensity, with the primary offset by $+0.6$
for clarity. The strong absorption lines include H$\epsilon$,
H$\delta$, H$\gamma$, and H$\beta$, and \ion{He}{i} $\lambda \lambda$
4026, 4144, 4388, 4471\AA.}
\label{disentangled_spec_fig}
\end{figure*}

\begin{table*}
\caption{Basic parameters for the targets studied here. Columns list
the stars OGLE identification (the first figure refers to the OGLE
field, the second the star number within that field); the 
co-ordinates; photometric period;  and OGLE $I$ magnitude and
colours outside eclipse (the $I$ magnitudes are quadrature values, returned 
from the normalization of the light-curve analysis described in Section~\ref{lca_sec}).}
\label{basic_param_tab}
\begin{tabular}{lccccccccc}
\hline
Star  & $\alpha$ (J2000) & $\delta$ (J2000) & $P_{\rm orb}$ & $I$ & $B-V$ & $V-I$ \\
Number&    h m s      &$^\circ$ $'$ $''$  &    d   &  \\ 
\hline
5 038089 & 00 49 01.85 & $-$73 06 06.9  & 2.38946 & 15.228 & $-0.222$ & $-0.144$  \\
5 202153 & 00 50 27.93 & $-$73 03 16.1  & 4.60677 & 14.265 & $-0.215$ & $-0.150$   \\ 
5 316725 & 00 51 05.95 & $-$72 40 56.7  & 2.55606 & 14.691 & $-0.257$ & $-0.142$   \\
6 077224 & 00 51 50.13 & $-$72 39 22.7  & 3.82087 & 14.009 & $-0.128$ & $-0.148$  \\
6 158118 & 00 52 19.28 & $-$72 41 51.7  & 2.57832 & 15.057 & $-0.085$ & $-0.156$   \\
6 215965 & 00 53 33.35 & $-$72 56 24.1  & 3.94604 & 14.123 & $-0.204$ & $-0.209$   \\
7 243913 & 00 56 56.34 & $-$72 49 06.4  & 2.63160 & 14.859 & $-0.150$ & $-0.096$   \\
9 175323 & 01 03 21.27 & $-$72 05 37.8  & 2.20596 & 13.675 & $-0.232$ & $-0.181$   \\
11 30116 & 01 06 24.86 & $-$72 12 48.3  & 2.95427 & 14.878 & $-0.200$ & $-0.167$   \\
11 57855 & 01 07 31.44 & $-$72 19 52.9  & 1.29695 & 16.000 & $-0.248$ & $-0.201$   \\
\hline
\end{tabular}
\end{table*}

\section{Data analysis}
\label{dis_sec}

The spectra were first normalized by using low-order polynomial fits
to interactively defined line-free continuum regions, and were then
velocity-corrected to the heliocentric frame.  Throughout, we assumed
circular orbits for all our objects, since their orbital periods are
short, the sizes of the stars relative to their
separation prove to be large ($r/a > 0.25$), and all the light curves have
secondary minima at orbital phase 0.5.

\subsection{Component spectra and orbital parameters}

We used two independent methods to obtain the velocity semi-amplitudes
and to separate the component spectra. The first was the disentangling
algorithm of Simon \& Sturm (1994), in which an over-determined matrix
equation is constructed from the data and a {\em given} orbital
solution, and is solved by using singular-value decomposition (SVD).
We obtained the best-fitting orbital solution by minimization of the
SVD residuals (by using a grid search followed by a downhill simplex).

The second method involves fitting for the two component spectra by
using a constrained non-linear least-squares algorithm (Byrd et al.\
1995), again for a given set of orbital parameters.  A genetic
algorithm (Charbonneau \& Knapp 1995) was used to search for the
orbital parameters which yielded the minimum in $\chi^2$ hyperspace.

Both methods assume that the component spectra do not change with
orbital phase.  Though not strictly correct (because of the
aspect-dependent effects of gravity darkening and `reflection'), this
is a good approximation, except during eclipses, when the apparent
intensity ratios of the components change significantly.  Spectra
obtained during eclipses were therefore excluded from the analyses.
This entails no loss of leverage on the determination of the velocity
semi-amplitudes, which are constrained by spectra obtained at
quadratures, but does mean that the time of maximum positive velocity
of the primary is less well-constrained. We therefore fixed the
spectroscopic phases using the OGLE ephemerides.

We found good agreement between the two RV analysis methods. In
practice we adopted the least-squares technique (which is computationally
faster) to disentangle the spectra after the orbital parameters had
been determined using the Simon \& Sturm approach.

The systems considered here may have significant tidal distortions, in
which case the centre of {\em light} (specifically, the weighted mean
line velocity) will not follow the centre of {\em mass} (see e.g.,
Kopal 1959, Hilditch 2001 for descriptions and bibliographic references).
We used the light-curve solutions (Section~\ref{lca_sec}) to determine
these so-called `non-Keplerian' corrections, and applied them to the
derived velocity semi-amplitudes.  We found that a single iteration
between the light-curve and radial-velocity solutions achieved
convergence in these corrections, which are typically on the order of
5\,\kms\ (and always less than 10\,\kms).  The final values for the
velocity semi-amplitudes for the primary\footnote{We use the
  convention that the primary is the star eclipsed at primary (i.e.,
  deeper) eclipse; it has the higher surface brightness, but is not
  necessarily the more massive component.} and secondary components
($K_1$ and $K_2$) are listed in Table~\ref{rv_tab}, together with the
formal uncertainties from the solutions. Further details of individual
solutions are included in Section~\ref{indiv_txt}.

Of course, the separated spectra of the binary components have
arbitrary normalizations: there is complete redundancy between line
strengths and relative continuum levels (i.e., it is, in principle,
impossible to distinguish between a relatively faint secondary with
intrinsically strong absorption lines, and a relatively bright
secondary with intrinsically weak absorption lines).  However, by
requiring the line strengths in the spectra of the two components to
be in the expected ratio, it is possible to obtain a spectroscopic
estimate of their relative ($\sim{B}$-band) continuum fluxes.

Systemic velocities for the primary and secondary components were
determined by cross-correlating the disentangled spectra with that of
the O9 star 10~Lac (HD~214680) which has a heliocentric radial
velocity of $-9$\,\kms (Holmgren, Hill \& Fisher 1990). The
cross-correlation function was fitted, by least squares, with a
Gaussian. We found that the systemic velocities of the components
in a given binary agreed to within their associated formal errors, and we quote the mean
systemic velocity for each system in Table~\ref{rv_tab}. The
velocities, which range from 164 to 211 \kms, are consistent with the
targets' location according to the radial-velocity maps given by
Mathewson, Ford \& Visvanathan (1988).

Finally, it should be noted that the orbital parameters are determined
directly from the ensemble of spectra of a given target, with no
intermediate explicit radial-velocity measurements. However, for the
convenience of future researchers, we have, {\em as a separate
  exercise}, measured the stellar radial velocities, by least-squares
fitting the disentangled component spectra to the observed spectra;
results are given in Appendix~A. Although traditional circular-orbit
spectroscopic solutions of these measured velocities yield
semi-amplitudes that agree, within the errors, with those listed in
Table~\ref{rv_tab}, the latter -- which include small non-Keplerian
corrections -- offer the best representations of the orbital
characteristics available at present.

\subsection{Spectroscopic classification}

The separate component spectra of all the stars were classified by
reference to the digital atlas of galactic OB stars by Walborn \&
Fitzpatrick (1990) and the study of SMC B~supergiants by Lennon
(1997).  (The lack of absolute normalizations for the component
spectra is of no consequence for spectroscopic classifications, which
depend only on relative line strengths.)  This task was performed
independently by all three authors in order to quantify the
uncertainities associated with the spectral types.  Typically, we
found that our classifications agreed to within one temperature type.
(We did not generally classify the luminosity class, since the
traditional indicators include metal lines, which are of course much weaker
in the SMC compared to the Galactic templates.) The adopted spectral
types, which represent the mode of the three individual estimates, are
included in Table~\ref{light_curve_table}.

\subsection{Light-curve analysis}

\label{lca_sec}

It is well known that light curves with alternating total and annular
eclipses (complete eclipses) provide the best determinations of the
photometric elements. Photometric determinations of the mass-ratio ($q$)
can approach the accuracy of determinations from independent
spectroscopic data (radial velocity curves of both components) but
only in such favoured cases.

In this particular sample of 10 binaries, selected principally by the
number and quality of spectra secured around both quadratures, 9 out
of 10 systems display only {\em partial} eclipses. Only 5~202153 has
total/annular eclipses but its light curve has notable asymmetries as
noted below. Accordingly, the possible separation of the dependence of
light-curve shape upon inclination ($i$) and mass ratio ($q$) is denied,
and it is again well established that a light curve may be fitted
equally well by a range of pairs of values $(i,q)$ that can be substantial,
dependent primarily upon the degree of distortion of both stars.

For these reasons we adopted a pragmatic procedure of solving each
light curve with the mass ratio fixed at the value determined
spectroscopically. This breaks the degeneracy between $i$ and $q$ and
in our experience yields better results than any grid search for a
minimum $\chi^{2}$ in $(i,q)$ space from the photometry alone. For
most systems, this procedure of adopting the spectroscopic $q$ has
enabled $I$-band light-curve solutions to be found readily that also
reproduce the flux ratio observed spectroscopically in the $B$-band.
In a few cases, noted below, it proved necessary to constrain the
solution further by using the spectroscopic $B$-band flux ratio to fix
either the secondary temperature or secondary radius.

In summary, we have used all the observational data available to us to
establish the set of parameters that matches best all of the data. In
a few cases, the light-curve solution is slightly poorer than would be
determined by concentrating on the light curve alone, but such
light-curve solutions do not agree with other parts of the available
data.  It should further be noted that by undertaking an analysis of a
large sample, problems with individual systems, and uncertainties
resulting from partial eclipses, are both diluted, and should average
out.

Fits were made to the Cousins $I$-band OGLE photometry of the targets,
using Hill's {\sc light2} code (Hill 1979; Hill \& Rucinski 1993).
The light-curve analysis yields, essentially, the surface-brightness
ratio of the components, although it offers no {\em direct}
information on the actual stellar temperatures.  We therefore fixed
the primary's temperature using a spectral-type--temperature
calibration (see Section~\ref{temp_txt}).  The mass ratio was also
fixed, from the radial-velocity solution, and the photometry was then
usually solved for the primary and secondary radii, the orbital
inclination, and the secondary temperature (which follows from the
surface-brightness ratio). Where preliminary solutions indicated
Roche-lobe-filling components, the corresponding radii were
subsequently constrained accordingly.  Results are given in
Table~\ref{light_curve_table}; model fits are compared to the data in
Figure~\ref{photo_fig}; and details of individual fits follow.  In all
cases, we assumed synchronous rotation (in the absence of evidence to
the contrary), limb-darkening coefficients appropriate to the local
temperatures, and gravity-darkening and `reflection' coefficients
appropriate to ionized radiative envelopes.

We quote ``surface area'' radii, viz
\begin{equation}
R^2  = \frac{1}{4 \pi}\int{dA} 
\end{equation}
and mean temperatures;
\begin{equation}
T_{\rm mean}^4 = L  / 4 \pi \sigma R^2
\end{equation}
where 
\begin{equation}
L = \sigma \int{ T^4}dA
\end{equation}
and $\sigma$ is the Stefan-Boltzmann constant. We note that the
differences between the ``mean area'' radius and the oft-quoted ``mean
volume'' radius are generally rather small (smaller in fact that the
errors on the radii). The errors quoted are taken from the covariance
matrix of the light-curve solution, and are formally propagated
through to the final system parameters.

Udalski et al. (1998b) make no explicit statement about the accuracy
of their light-curve ephemerides. We found it necessary to allow for
shifts in orbital phase of an entire light curve relative to a
theoretical model fit in order to minimize the $\chi^2$ value. The
largest phase shift was $+0.0100$ for the system 6~77224, with the
remainder all less than $\pm0.0060$, and some being zero. The average
phase shift for the set of 10 systems was $+0.0015\pm\,0.0046$. For
completeness, these ephemeris revisions were incorporated in the
disentangling procedures described in Section~\ref{dis_sec}.

\begin{table}
\caption{Radial-velocity solutions. Listed are the OGLE designation,
  the velocity semi-amplitudes  of the primary and
  secondary stars ($K_1$ and $K_2$ respectively), 
and the systemic velocity, $V_0$.}
\label{rv_tab}
\begin{center}
\begin{tabular}{lccc}
\hline
\multicolumn{1}{c}{Star}
& $K_1$  & $K_2$  & $ V_0$\\
\multicolumn{1}{c}{Number}
& (\kms) & (\kms) & (\kms) \\
\hline
5 038089 & $271 \pm 10         $  & $242 \pm 10         $ & $189 \pm 8$\\
5 202153 & $157 \pm\phantom{0}8$  & $250 \pm\phantom{0}4$ & $172 \pm 5$\\
5 316725 & $154 \pm 11         $  & $295 \pm\phantom{0}9$ & $169 \pm 9$\\
6 077224 & $165 \pm\phantom{0}9$  & $201 \pm\phantom{0}4$ & $178 \pm 5$\\
6 158118 & $138 \pm\phantom{0}5$  & $280 \pm\phantom{0}4$ & $188 \pm 8$\\
6 215965 & $217 \pm\phantom{0}5$  & $202 \pm\phantom{0}2$ & $164 \pm 6$\\
7 243913 & $164 \pm\phantom{0}8$  & $290 \pm\phantom{0}6$ & $197 \pm 4$\\
9 175323 & $193 \pm\phantom{0}9$  & $280 \pm\phantom{0}7$ & $211 \pm 3$\\
11 30116 & $144 \pm 21         $  & $269 \pm 10         $ & $201 \pm 4$\\
11 57855 & $194 \pm 16         $  & $293 \pm\phantom{0}8$ & $168 \pm 9$\\
\hline
\end{tabular}
\end{center}
\end{table}

\subsection{Notes on individual targets}
\label{indiv_txt}
\subsubsection{5 038089}
\label{5_038089_sec}

This target was in the overlap region of the 2dF fields.  Of the 32
spectra (typical S:N of 40 per pixel), 13 were within $\pm0.1$ in
orbital phase of the quadratures. Both components are clearly visible
in the spectra. The flux ratio in the blue, as estimated from the
component spectra, is close to unity.

The $I$-band light curve shows 0.3-mag-deep eclipses, of approximately
equal depth, while ellipsoidal variation is evident between eclipses,
indicating significant tidal distortions (Figure~\ref{photo_fig}a).  A
detached configuration was found for this system.  The solution
predicts nearly-equal blue-region fluxes, which is consistent with the
spectroscopy. The (O-C) curve is flat through all orbital phases
demonstrating a very good fit to the light curve.

\subsubsection{5 202153}
\label{5_202153_sec}

This star is one of brightest considered here, and the quality of its
spectra (S:N $\sim$50) reflects this.  Fifteen spectra were used in
the disentangling procedure, but the second quadrature is less well
covered than the first. The primary spectrum shows strong
\ion{He}{ii} absorption, which is undetected in the secondary
spectrum.

We found that the light-curve solution rapidly converged to a
configuration in which the secondary fills its Roche lobe. We
therefore modeled the light-curve with a semi-detached configuration,
solving for the secondary temperature, the primary radius, and the
inclination. The solution indicates stars of roughly equal luminosity
in $B$ (which is consistent with the relative strengths of the lines
in the disentangled spectra).

The $I$-band light curve for this star is clearly asymmetric, with the
maximum at first quadrature being 0.03m brighter than that at
second quadrature. Additionally, just before ingress into primary
eclipse (around orbital phase 0.9) the observed curve is lower by up
to 0.06m than the corresponding egress phases around phase
0.1.  These asymmetries are clearly seen in Fig.~\ref{photo_fig}b
which compares the observations with the best-fit model light curve
that is mirror-symmetric about phase 0.5. The asymmetry around ingress
is suggestive of attenuation by an accretion stream from the
Roche-lobe filling secondary falling towards the primary, but the
explanation for the higher maximum at first quadrature is unclear. It
is difficult to quantify how much these asymmetries in the $I$-band
light curve affect the determination of the system parameters and its
distance modulus.

\subsubsection{5 316725}
\label{5_316725_sec}

Our spectra typically have S:N ratios of 30--40. The quadrature phases
are well-sampled, with 6 observations at first quadrature and 8 at
second. The secondary's spectral features appear to be 
stronger than those of the primary.

The $I$-band light curve shows relatively deep eclipses of comparable
depth ($\sim{0.4}$ mag), with significant ellipsoidal variations.  We
found that the solution consistently converged to a semi-detached
configuration, and we therefore solved for the primary radius, the
secondary temperature and the inclination, with the secondary filling
its Roche lobe. The solution gives a $B$-band flux ratio (p/s) of
0.7--0.8 around the quadratures, which is consistent with the relative
line strengths in the disentangled spectra. The (O-C) curve is flat
through all orbital phases demonstrating a very good fit to the light
curve.

\subsubsection{6 077224}
\label{6_077224_sec}

The spectra of this binary typically have a S:N of $\ga$50, and
quadrature phases are well sampled (11 at first, 9 at second).  The
primary's spectrum shows \ion{He}{ii} features that are undetectable
in the secondary, whose absorption lines appear otherwise to be
stronger than the primary, indicating an inverted light ratio in this
object.

The $I$-band light curve shows large ellipsoidal variation combined
with shallow eclipses of approximately equal depth. A semi-detached
solution gives the best fit, and we solved for the primary radius,
secondary temperature, and the inclination (with the secondary star
filling its Roche lobe).  The light-curve solution gives a
$B$-band flux ratio (p/s) of 0.8, in accord with the line strengths of
the disentangled component spectra. The (O-C) curve is nearly flat
through all orbital phases with only a small offset just before
ingress to primary eclipse in this semi-detached system.

\subsubsection{6 158118}
\label{6_158118_sec}

This is one of the fainter objects in the sample ($V \simeq 15.1$) and
the spectroscopic S:N is relatively poor (typically $\sim$30).
Quadrature phases are reasonably well covered, with 3 spectra at first
quadrature and 9 at second.  We find that the primary's spectrum is
much stronger than that of the secondary.

Although the phase coverage of the light curve during primary eclipse
is not very good, the light curve is that of a typical EB type.
Initially, we solved for the inclination, the secondary temperature,
and the primary radius, fixing the secondary radius at the Roche-lobe
value.  However, we found that the light ratio of the solution was
inconsistent with the spectroscopy, in the sense that the photometric
ratio was much greater than that suggested by the spectroscopy.  From
the equivalent widths of the strongest absorption lines in the
disentangled spectra we estimate a flux ratio of $\sim$2.5 at
$\sim4400$\AA.  By fixing the secondary temperature (i.e., brightness)
according to this light ratio we were able to obtain a satisfactory
and consistent fit to the light curve, and solving only for the
primary radius and the inclination. Secondary eclipse is fitted very
well, but primary eclipse shows some asymmetry with the phases just
before ingress being 0.02~mag fainter than egress.

\subsubsection{6 215965}
\label{6_215965_sec}

The light curve of this object has been analysed by Wyithe \& Wilson
(2001), who assumed a mass ratio of unity.  They suggested that the
system should be a good distance indicator on the grounds that it may show
complete eclipses.

The S:N of our spectral observations of this target vary between 20
and 100 per pixel, with a median of about 50. The quadrature phases
are well sampled, and features of both components are clearly seen in
spectra obtained at these phases. The separated component spectra
indicate a light ratio (p/s) of around 0.9.

\begin{figure*}
\includegraphics{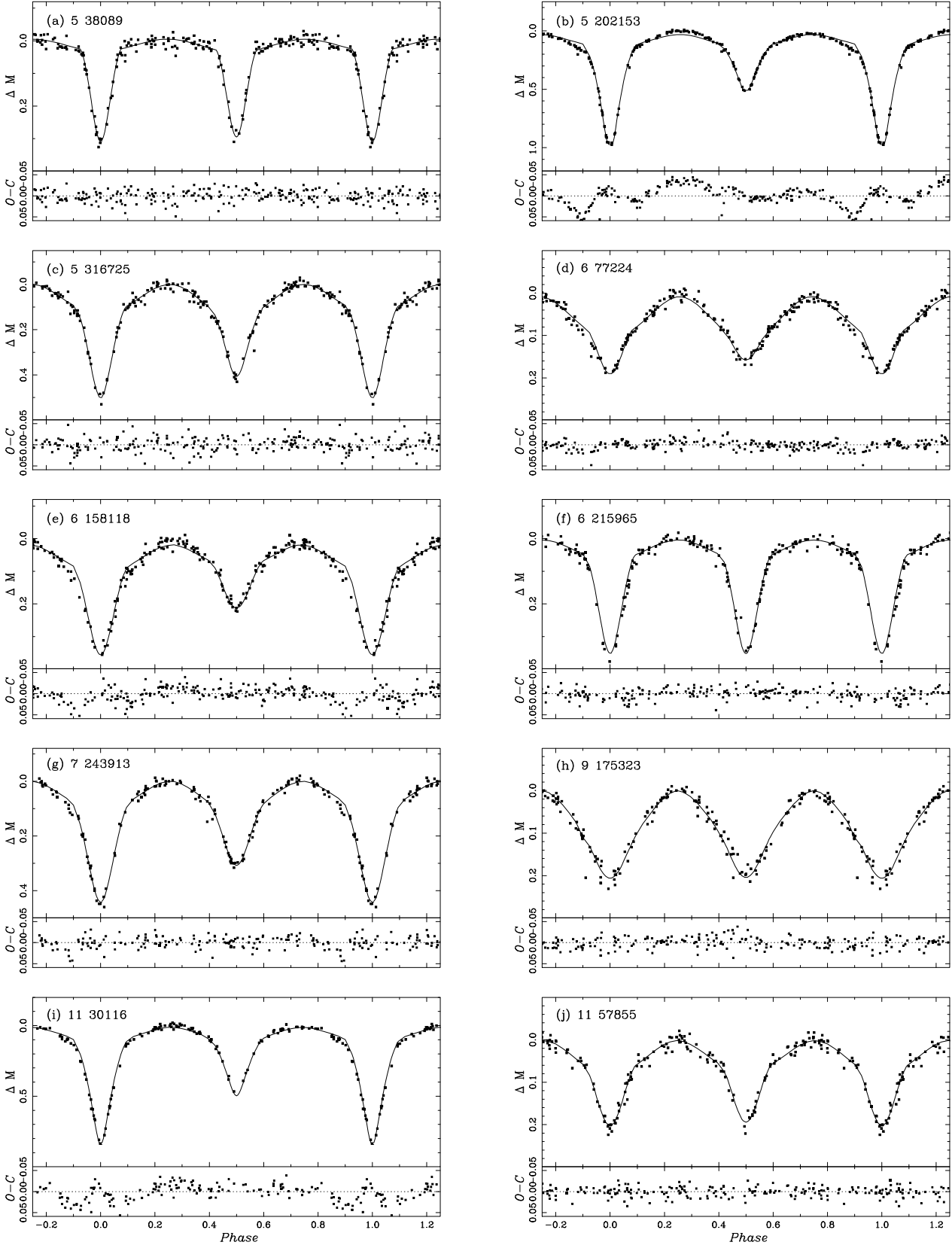}
\caption{$I$-band light-curve fits for the sample, plotted over 1.5
  cycles. Each panel shows the data (points) and fit (solid) line
  (upper subpanel), along with the $(O-C)$ scatter about the fit
  (lower subpanel).  Note that the phases plotted here are {\em
    photometric}.}
\label{photo_fig}
\end{figure*}

The light curve shows relatively deep eclipses ($\sim{0.4}$~mag, with
the secondary eclipse about 0.05~mag shallower -- see
Figure~\ref{photo_fig}f). The entry ingress of the primary eclipse is
quite poorly sampled, although the light-curve itself is of relatively
high quality. There is considerable variability outside eclipses,
indicating that the stars are significantly tidally distorted. We
fitted for the inclination, the two relative radii, and the secondary
temperature.  This solution fitted the $I$-band light curve well, but
returned a $B$-band flux ratio (p/s) of 1.3, which conflicts with the
spectroscopic evidence of ~0.9. It is most likely that the light curve
solution has been compromised by the less well sampled sections of the
eclipse curves. Therefore, we fixed the B-band flux ratio at 0.9,
adopted a temperature for the secondary of 27800\,K corresponding to
its spectral type of B0.5, and calculated the radius of the secondary
to provide the observed B-band flux ratio. Light-curve solutions for
the primary radius and the inclination, with the secondary radius
recalculated at each iteration then provided a final solution
(Table~\ref{light_curve_table}) of the light curve that agrees with
the spectroscopic data, and fits the light curve almost as well as the
first solution for 4 parameters. The whole light curve is fitted reasonably
well, particularly secondary eclipse, except for a
$+0.007\pm0.012$(sd) mean (O-C) offset for the data within primary
eclipse. The eclipses are partial in this solution.

Our light-curve solution is not consistent with that published by Wyithe
\& Wilson (2001), which (in the absence of spectroscopic data) was
performed without a constraint on the light ratio.

\begin{table*}
\caption{Light-curve solutions. Columns give the OGLE target
  designation;  the nature of the system according to the adopted solution (detached,
semi-detached, or contact); and then, for the primary and secondary stars,
our spectral types;  the (flux-averaged)
effective temperatures; the mean stellar radii relative to
  the binary separation; the inclination; the component masses; the
  absolute mean radii; and the equivalent surface gravities
$(GM/\overline{R}^2)$.
The primaries' temperatures are adopted on the basis of the adopted
spectral types (cf.\ Section~\ref{temp_txt}), and the secondary
temperatures follow from the radii and primary:secondary intensity
ratios 
at quadratures (Section~\ref{metho_txt}).}
\label{light_curve_table}
\begin{tabular}{lcclllcccc}
\hline

\multicolumn{1}{c}{Star}     &System&    &\multicolumn{1}{c}{Spectral}  & \multicolumn{1}{c}{Temperature} & \multicolumn{1}{c}{Relative} & Inclination & Mass  & Radius   & $\log g$  \\
\multicolumn{1}{c}{Number}   &type&    &\multicolumn{1}{c}{type}   & \multicolumn{1}{c}{(K)}      &  \multicolumn{1}{c}{radius}  & ($^\circ$)    &(\msol)& (\rsol)& dex (cgs) \\
\hline
5 038089  &d& p  & B0        & 30100 (fixed)    & $0.252 \pm 0.005$ & $76.9 \pm 0.3$ & $17.1 \pm 1.5$ &  $\phantom{0}6.1 \pm 0.2$ & $4.10 \pm 0.05$ \\
           && s  & B0.2      & 29180 (fixed)    & $0.249 \pm 0.008$ &                & $19.1 \pm 1.6$ &  $\phantom{0}6.1 \pm 0.3$ & $4.15 \pm 0.05$ \\
\rule[0mm]{0mm}{3.5mm}
5 202153 &s& p  & O9.5      & 32200 (fixed)    & $0.257 \pm 0.004$ & $87.0 \pm 0.2$ & $19.9 \pm 1.1$ &  $\phantom{0}9.5 \pm 0.3$ & $3.78 \pm 0.03$ \\
          && s  & B0.5      & $23450 \pm 200$  & 0.346 (fixed)      &                & $12.5 \pm 1.2$ &  $          12.8 \pm 0.3$ & $3.32 \pm 0.04$ \\
\rule[0mm]{0mm}{3.5mm}
5 316725 &s& p  & O9        & 33800 (fixed)    & $0.264 \pm 0.007$ & $77.6 \pm 0.3$ & $17.0 \pm 1.5$ &  $\phantom{0}6.1 \pm 0.3$ & $4.09 \pm 0.05$ \\
          && s  & O9        & $31540 \pm 170$  & 0.330 (fixed)      &                & $\phantom{0}8.9 \pm 1.1$ &  $\phantom{0}7.7 \pm 0.2$ & $3.62 \pm 0.06$ \\
\rule[0mm]{0mm}{3.5mm}
6 077224  &s& p  & O9.5      & 32200 (fixed)    & $0.283 \pm 0.003$ & $61.1 \pm 0.2$ & $15.9 \pm 1.0$ &  $\phantom{0}8.9 \pm 0.3$ & $3.74 \pm 0.04$ \\
           && s  & B0.5      & $25710 \pm  530$  & 0.365 (fixed)      &                & $13.1 \pm 1.4$ &  $11.5 \pm 0.3$ & $3.43 \pm 0.05$ \\
\rule[0mm]{0mm}{3.5mm}
6 158118 &s& p  & B0.2      & 29180 (fixed)    & $0.334 \pm 0.004$ & $69.2 \pm 0.2$ & $16.0 \pm 0.7$ &  $\phantom{0}7.6 \pm 0.2$ & $3.88 \pm 0.02$ \\
          && s  & B1        & 18380 (fixed)    & 0.324 (fixed)      &                & $\phantom{0}7.9 \pm 0.5$ &  $\phantom{0}7.4 \pm 0.1$ & $3.60 \pm 0.03$ \\
%
%
\rule[0mm]{0mm}{3.5mm}
6 215965 &d& p  & B0.5      & 27800 (fixed)    & $0.294 \pm 0.004$ &$75.6 \pm 0.3$ & $16.0 \pm 0.5$ & $\phantom{0}9.9 \pm 0.2$ & $3.66 \pm 0.02$ \\
          && s  & B0.5      & 27800 (fixed)    &  0.307 (fixed)     &               & $17.2 \pm 0.8$ &           $10.4 \pm 0.1$ & $3.65 \pm 0.03$ \\

%

\rule[0mm]{0mm}{3.5mm}
7 243913 &s& p  & O9.5      & 32200 (fixed)    & $0.334 \pm 0.003$ & $73.3 \pm 0.2$ & $18.6 \pm 1.1$ &  $\phantom{0}8.2 \pm 0.2$ & $3.88 \pm 0.03$ \\
          && s  & B1.5      & $25960\pm 310$   &  0.336 (fixed)     &                & $10.5 \pm 0.9$ &  $\phantom{0}8.3 \pm 0.2$ & $3.62 \pm 0.04$ \\
\rule[0mm]{0mm}{3.5mm}
9 175323 &c& p  & O6.5      & 39250 (fixed)    &  0.418   (fixed)     & $58.0 \pm 0.2$ & $23.6 \pm 1.6$ &  $10.2           \pm 0.3$ & $3.79 \pm 0.04$ \\
          && s  & O7        & 38500 (fixed)    &  0.349   (fixed)     &                & $16.2 \pm 1.5$ &  $\phantom{0}8.5 \pm 0.2$ & $3.79 \pm 0.04$ \\
\rule[0mm]{0mm}{3.5mm}
11 30116 &s& p  & B0.5      & 27800 (fixed)    & $0.306 \pm 0.003$ & $83.8 \pm 0.2$ & $14.3 \pm 1.9$ &  $\phantom{0}7.4 \pm 0.4$ & $3.85 \pm 0.08$ \\
          && s  & B2        & 20340 (fixed)    &  0.332  (fixed)    &                & $\phantom{0}7.7 \pm 1.9$ & $\phantom{0}8.1 \pm 0.5$ & $3.51 \pm 0.12$  \\
\rule[0mm]{0mm}{3.5mm}
11 57855 &d& p  & B0.5      & 27800 (fixed)    & $0.381 \pm 0.006$ & $65.6 \pm 0.7$ &           $12.4 \pm 1.1$ &  $\phantom{0}5.2 \pm 0.2$ & $4.09 \pm 0.05$ \\
          && s  & B1        & $25970\pm  930$  & $0.272 \pm 0.014$ &                & $\phantom{0}8.2 \pm 1.3$ &  $\phantom{0}3.7 \pm 0.2$ & $4.21 \pm 0.09$ \\
\hline
\end{tabular}
\end{table*}

\subsubsection{7 243913}
\label{7_243913_sec}

This system has $V=14.8$, and the spectra obtained have a S:N between
20 and 40. The coverage is good at second quadrature (9 spectra), but
is relatively poor at first quadrature (3 spectra only).  The
secondary's spectral features are ~0.5 mag weaker than those of the
primary.

The $I$-band light curve may be fitted equally well at a fixed
spectroscopic mass ratio of 1.77 by either a detached configuration
with the secondary cooler and smaller than the primary, or by a
semi-detached configuration with the secondary filling its Roche lobe
at a radius comparable with that of the primary and at a higher
temperature than in the detached case. However, the detached
configuration suggests a $B$-band flux ratio of greater than 5,
whereas the semi-detached configuration indicates a B-band flux ratio
of 1.4, in excellent agreement with the spectroscopic data. This is a
well-fitted light curve (a flat (O-C) curve) except for a short
section of disparity just before ingress to primary eclipse.

\subsubsection{9 175323}
\label{9_175323_sec}

At $V=13.7$ this is the brightest target in the present sample, and
the spectroscopic data are of correspondingly good quality (S:N
typically $\ga$50). Although the object was not in the overlap region
of the 2dF fields, and hence there are currently only 16 phase points
available, the orbit was fortuitously phased for our run, and 11 of
the spectra were obtained near quadrature phases. The presence of very
strong \ion{He}{ii} absorption and \ion{N}{iii} $\lambda$4634--40--42
emission indicates very luminous components, and we estimate spectral
types of O6.5 and O7 for the primary and secondary respectively.

The $I$-band light curve strongly suggests a contact system, and
attempts to secure solutions with a fill-out factor exceeding the
Roche-lobe sizes were made, as for standard contact binaries. These
solutions failed, with the code restoring the fill-out factor back to
marginally separated stars.  Accordingly, a semi-detached
configuration was attempted with the $B$-band flux ratio of 1.3 being
adopted from the spectroscopy to fix the secondary temperature.
Solutions were made for the primary radius and the orbital
inclination. We find that the primary component very nearly fills its
Roche lobe (a filling factor of 99 per cent) and that the system is
essentially in a marginal contact state. The (O-C) curve is flat
through all orbital phases demonstrating a very good fit to the light
curve. We note that the solution is very similar to two other
well-studied systems, V348 Car (Hilditch \& Lloyd-Evans 1985) and
SX~Aur (Bell, Adamson \& Hilditch 1987).

\subsubsection{11 30116}
\label{11_30116_sec}

11~30116 has $V$ magnitude of 14.7, and a S:N of 30 is typical of our
spectroscopic data of this system. The star is not in the 2dF overlap
region, although 13 of the 16 spectra were obtained near quadrature
phases. The primary's spectral features are stronger than those of the
secondary.

The $I$-band light curve is of the EB type, but the phase sampling of
the photometry is poor during the middle of the secondary eclipse.
From an initially detached description, the solutions converged rapidly
to a semi-detached configuration. We therefore fixed the secondary
radius at the Roche-lobe value, and at first solved for the secondary
temperature, the inclination, and the primary radius. However, the
solution gave a $B$-band light ratio of 2.3, which is inconsistent
with the relative strength of the spectral lines. We therefore
estimated the spectroscopic light ratio from the strengths of the
strongest H/He lines, obtaining a value of 1.5 (p/s).  The solution
given in Table~\ref{light_curve_table} was obtained by fixing the
$B$-band light ratio at this value, and solving only for the primary
radius and the inclination. Most parts of the light curve are fitted
very well with some disparity before ingress to primary eclipse and
through egress.

\subsubsection{11 57855}
\label{11_57855_sec}

This short-period system is the faintest of the ten considered here,
and this is reflected in the quality of the spectra, which have a
typical S:N of $\sim$20--30. The target is located outside the overlap
region, in Field~2, and therefore has only 16 phase points in total,
of which 7 were close to quadrature phases.  We solved the light curve
for the secondary temperature, the inclination and the two radii. The
solution indicates a detached system, with the primary close to
filling its Roche Lobe (92\% by volume). This light-curve solution
predicts a $B$-band flux ratio of 2.4 around the quadratures, in good
agreement with the estimates from the spectroscopy of 2.0-2.5.  The
(O-C) curve is flat through all orbital phases demonstrating a very
good fit to the light curve.

\begin{figure*}
\includegraphics{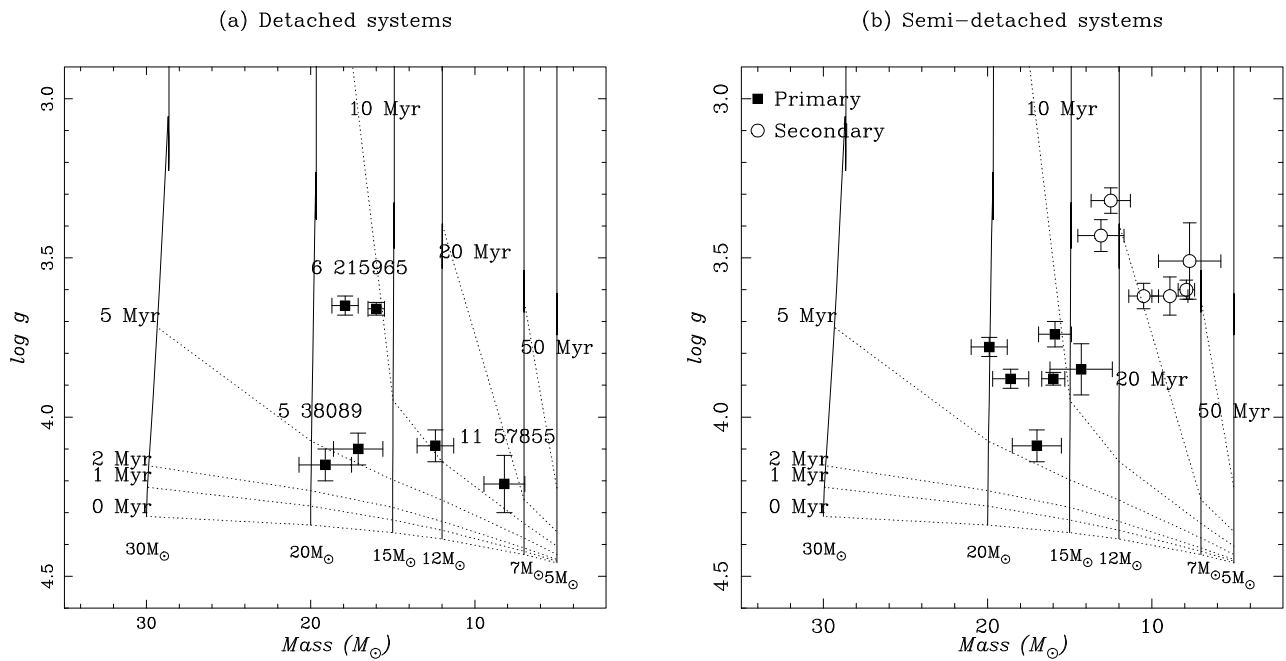}
\caption{The parameters of (a) the detached and (b) the semi-detached
  systems, plotted in the mass--$\log g$ plane. For the detached
  systems the components are plotted as filled squares, while for the
  semi-detached systems the primaries are plotted as filled squares
  and the secondaries as open circles. Evolutionary tracks
  for different masses  ($Z=0.004$; Girardi et al.\ 2000) are plotted
   for different masses (solid lines), along with isochrones for 0, 1,
  2, 5, and 10\,Myr (dotted lines).}
\label{mass_logg_fig}
\end{figure*}

\section{Evolutionary characteristics}

\subsection{Detached systems}

Since our sample is limited to $P_{\rm orb} < 5$d, it is unsurprising
that only 3 of the 10 systems are detached. The B0+B0.2 system 5~38089
has a period of 2.4\,d and component masses of $17.1\pm1.5$\msol and
$19.1\pm1.6$\msol. Both stars are well within their Roche lobes, with
volume-filling factors of 69\% and 64\% for the primary and secondary,
respectively.

The longer-period system 6~215965 ($P_{\rm orb} = 3.9$d) contains two
B0.5 stars, and is both slightly more evolved (see
Figure~\ref{mass_logg_fig}) and of similar mass ($16.0\pm0.5$\msol for
the primary and $17.2\pm0.8$ for the secondary). Since its mass ratio
is close to unity this system does not critically test the evolution
models.

The short-period system 11~57855 contains an $12.4\pm1.1$-\msol, B0.5
primary and a $8.2\pm1.3$-\msol, B1 secondary, and this larger mass
ratio means this system offers a better test of the models.

We plot these systems in the mass--surface-gravity plane in
Figure~\ref{mass_logg_fig}(a), along with the $Z=0.004$ evolutionary
tracks by Girardi et al.\ (2000). It appears that 6~215965 is
approximately 10\,Myr old, whereas 5~038089 is younger (around
5\,Myr). Both components of 11~57855 lie on the 10\,Myr isochrone,
although the gravity of the secondary is quite poorly constrained (due
to a relatively large error on the relative radius of the secondary
from the light-curve fit). Analysis of further systems should reveal
more detached binaries with $q>1$, which may be used to test
critically the models.

\subsection{Semi-detached systems}

Six of the ten systems presented here are in a semi-detached
configuration in which the secondary star fills its Roche lobe. We
believe that these systems evolved to this state via case-A mass
transfer, in which the mass donor (now the less massive star) reached
its Roche lobe whilst on the main sequence, and that the stars are now
in the slow phase of case-A mass transfer (Wellstein, Langer \& Braun
2001). As expected, the secondary (mass-donor) stars are now
overluminous for their spectral types, while the primary stars (the
mass gainers) appear as slightly evolved main-sequence objects (see
Figure~\ref{mass_logg_fig}b).

The mass donors typically have masses of 6--8\msol, which is in good
agreement with the masses found for galactic semi-detached OB-star
binaries (see, for example, Figure~9 of Harries et al. 1998).

\section{The distance to the SMC: preamble}

In order to reveal clearly the central arguments which lead to the
distance estimates for our sample, and potential sources of error and
uncertainty, we first review the key steps.

\subsection{Light-curve analysis}

The quantities of interest which are obtained from the light-curve
analysis are the stellar radii, scaled to the size of the orbit; the
orbital inclination; and, essentially, the ratio of surface
brightnesses of the two components.  (In the absence of gravity
darkening, limb darkening, tidal distortions and reflection, the ratio of eclipse
depths gives the surface-brightness ratio directly.)  This ratio is
conveniently expressed in terms of stellar temperatures, but it is
important to recall that, for practical purposes, a light-curve
obtained in a single colour carries no direct information on the
actual temperatures.  The corollary to this is that the dimensional
information obtained from the light-curves (and our discussion to this
point) are essentially independent of the adopted temperature scale.

The light-curve solutions conducted here use well-established physics
encoded in mature software.   It is reasonable to suppose that the
light-curve analysis is not a source of significant systematic errors.

\subsection{Radial-velocity analysis}

The radial-velocity solution, together with the inclination from
the light-curve analysis, yields the size of the orbit (and hence
places the stellar radii on an absolute scale), and the stellar
masses.   Although the measurement techniques adopted herein have
been developed relatively recently, they seem robust, and hence it is
again reasonable to suppose that the radial-velocity orbits are not
a source of significant systematic errors.

\subsection{Absolute magnitudes and extinction}

Given the radius of a star, a knowledge of its surface brightness in
some photometric passband allows its absolute magnitude in that
passband to be calculated.  The apparent magnitude, corrected for
interstellar extinction, then gives the distance.  The calculation of
surface brightness and the determination of the extinction are crucial
factors (and important sources of potential systematic errors) and are
discussed in detail in the following sections. (Of course, the
temperature and surface brightness vary across the surface of a star
in a binary system, and we take proper account of that in our
calculations.)

\section{Absolute-magnitude calculations}

To determine the stellar surface brightness requires the assignment of
an effective temperature, and the calculation of a model atmosphere to
give the emergent flux as a function of wavelength.\footnote{It may
appear that a short-cut can be achieved by using the bolometric
correction.  That is, given the stellar radius and temperature, one
has the luminosity, hence $M_{\rm BOL}$; the effective temperature
also yields the bolometric correction, hence $M(V)$. However, the
bolometric correction is, in effect, a parameterization of the
$V$-band surface brightness, and is strongly model dependent for stars
such as those considered here, where the greater part of the emitted
flux emerges at unobservable wavelengths.}  For stars somewhat cooler
than those considered here, it may be possible to make a direct
estimate of the stellar temperatures from observed continuum fluxes
(e.g., Guinan et al.\ 1998), although even this approach entails
significant uncertainties (cf.\ Groenewegen \& Salaris 2001).  In our
case, we have to rely on a simple calibration of temperature as a
function of spectral type.

\subsection{Temperature scales}
\label{temp_txt}

\begin{table*}
\caption{Selected spectral-type--$T_{\rm eff}$ calibrations for late-O
and early-B dwarfs in units of kK.  Sources are Conti (1973; nLTE H/He models);
B\"ohm-Vitense (1981; semi-direct, and nLTE H/He); Underhill (1982;
semi-direct); Schmidt-Kaler (1982; semi-direct, and nLTE H/He);
Howarth \& Prinja (1989; nLTE H/He); Vacca, Garmany \& Shull (1996;
nLTE H/He); Crowther (1997, updated; various); Martins et al.\ (2002;
unblanketed [u] and line-blanketed [$\ell$] nLTE models).  Not all
scales use independent primary data -- e.g., the BV81 and SK82 scales
both incorporate the C73 results.  The adopted SMC scale is discussed in
Section~\ref{temp_txt}}
\label{spteff_tab}
\begin{tabular}{lcccccccccccl}
\hline
    & C73  & BV81&  U82& SK82& HP89& VGS & PAC & MSH{\em u}& MSH$\ell$ &$\quad$ &Adopted&     \\
O6  & 42.0 & 39.0&     & 41.0& 42.0& 43.6& 42.0& 42.0& 39.5 &&40.0&    O6  \\
O6.5& 40.0 &     &     &     & 40.5& 42.3&     & 40.5& 38.2 &&    &    O6.5\\
O7  & 38.5 & 37.5&     & 38.0& 39.0& 41.0& 40.0& 39.5& 37.0 &&38.5&    O7  \\
O7.5& 37.5 &     &     &     & 37.5& 39.7&     & 38.5& 36.0 &&    &    O7.5\\
O8  & 36.5 & 35.6&     & 35.8& 36.5& 38.4& 35.0& 37.0& 35.2 &&36.3&    O8  \\
O8.5& 35.5 & 34.6&     &     & 35.0& 37.2&     & 36.2& 34.5 &&35.3&    O8.5\\
O9  & 34.5 & 33.2&     & 33.0& 34.0& 35.9& 31.8& 35.5& 34.0 &&33.8&    O9  \\
O9.5& 33.0 & 31.6&     &     & 33.0& 34.6& 30.8& 34.5& 33.0 &&32.2&    O9.5\\
B0  &      & 29.5& 30.8& 30.0&     & 33.3& 29.8&     &      &&30.1&    B0  \\
B0.5&      & 27.3& 29.3&     &     & 32.1& 28.5&     &      &&27.8&    B0.5\\
B1  &      & 25.0& 26.9& 25.4&     &     & 26.1&     &      &&25.5&    B1  \\
B1.5&      & 23.0& 25.7&     &     &     & 24.8&     &      &&23.5&    B1.5\\
B2  &      & 21.5& 22.8& 22.0&     &     & 21.3&     &      &&21.9&    B2  \\

\hline
\end{tabular}
\end{table*}

Because much of the radiation of O- and early~B-type stars is emitted
in unobservable spectral regions,\footnote{More than 90\%\ of the flux
  of a 30kK, $\log{g} = 4.0$ SMC-metallicity model emerges at
  wavelengths shortwards of 300nm.} and because the slope of the
optical continuum is insensitive to temperature for these stars,
spectroscopic $T_{\rm eff}$ determinations must rely heavily on
modeling of the line spectra (in practice, He$\;${\sc i}/He$\;${\sc
  ii} line ratios).  `Traditional' temperature scales, exemplified by
those given by Conti (1973) and Howarth \& Prinja (1989), are based on
relatively simple non-LTE H/He models.  Subsequent work using
line-blanketed nLTE models (e.g., Hubeny, Heap \& Lanz 1998) and the
independent technique of comparison of fundamental stellar parameters
with evolutionary models (Hilditch et al. 1996; Harries \& Hilditch
1998) suggest that the simple models lead to temperatures which are
systematically too high, at a level of $\sim$1kK.

Table~\ref{spteff_tab} summarises a number of temperature scales that
have been published in the `nLTE era'.  In spite of the difficulties
alluded to above, there is a reasonable degree of agreement for
main-sequence stars, results being bracketed by the Vacca, Garmany \&
Shull (1996) scale, which is the hottest at all spectral types, and
the B\"ohm-Vitense (1981) scale, which is generally the coolest.  We
require a calibration that covers both late-O and early-B subtypes,
and adopt as a starting point the results given by B\"ohm-Vitense (1981).  

This is a rather venerable scale which, at these temperatures, is
derived in large part from studies by Conti (1973) and by Underhill et
al.\ (1979), which in turn are not above criticism (being based on
unblanketed models, and continuum-fitting techniques, respectively).
Nonetheless, the B\"ohm-Vitense scale is in good agreement with more
recent analyses (at least for near-main-sequence stars), and, in
particular, by adopting a relatively `cool' scale we acknowledge
the trends indicated by the binary-star and line-blanketed work.
Moreover, across the O9--B1 range, where most (80\%) of our stars
occur, the temperature span of the B\"ohm-Vitense scale, 8.2kK, is
greater than that of other scales.  We present an argument in
Section~\ref{tr_txt} that this is a desirable characteristic.

The B\"ohm-Vitense calibration (in common with all others) relates the
temperatures of {\em Galactic}\/ stars to their spectral types.  Even
though the spectral types of our stars are largely based on their
helium spectra, and hence are effectively independent of metallicity,
it would not be surprising if given helium-line ratios were obtained
at different effective temperatures for Galactic and SMC
metallicities.  Unblanketed models offer no direct insight into the
magnitude of such differences, but Martins, Schaerer \& Hillier (2002)
report that temperature differences between unblanketed models and
line-blanketed models with SMC metallicity are $\sim$60\%\ of those
between unblanketed and Galactic models.

In adopting an SMC temperature scale, we therefore increase the
B\"ohm-Vitense temperatures by 40\%\ of the temperature differences
implied by the Martins et al.\ unblanketed and (Galactic-metallicity)
blanketed models.  For late-O stars this leads to an increase in
temperature of $\sim$0.6kK, or $\sim$2\%\ of $T_{\rm eff}$, a scale
factor which we also apply to the B-star scale.  We will consider the
consequences of the adopted scale further when discussing the error
budget.

\subsection{Surface brightness}
\begin{table}
\caption{Surface-brightness characteristics of model atmospheres.
Models are identified by origin and $\log{g}$ (cgs, $\times$10): 
A ({\sc atlas9} solar-metallicity line-blanketed models), 
S (Howarth \& Lynas-Gray 1989, SMC-metallicity 
  line-blanketed {\sc atlas6} models),
T (Smith \& Howarth 1998, {\sc tlusty} nLTE H/He models), 
and BB (black-body).}
\label{surface_tab}
\begin{tabular}{lcc}
\hline
Model &
$\partial{\ln{S}}/\partial{T_{\rm eff}}$ &
$S(I)$ at 30kK\\
     & 
[kK$^{-1}$] & 
[10$^7$ erg cm$^{-2}$ s$^{-1}$ \AA$^{-1}$]\\
\hline
  A35& $6.8  \times 10^{-2} $&11.8  \\
  A40& $6.5  \times 10^{-2} $&11.1  \\
  S35& $6.7  \times 10^{-2} $&11.0  \\
  S40& $6.5  \times 10^{-2} $&10.4  \\
  T40& $6.1  \times 10^{-2} $& 9.9  \\
  BB & $4.5  \times 10^{-2} $&14.0  \\
\hline
\end{tabular}
\end{table}

We have examined the surface brightness, $S$, as a function of
temperature over the range $25\mbox{--}35$kK by using a grid of {\sc atlas9}
line-blanketed, solar-metallicity, LTE models (Kurucz 1993); the
Howarth \& Lynas-Gray (1989) line-blanketed, $\sim$SMC-metallicity,
LTE models; and the Smith \& Howarth (1998) nLTE H/He models.
The models were convolved with the photometric passbands given by
Bessell (1990).

For all models, we find $\partial{\ln{S}}/\partial{T_{\rm eff}} =
6\mbox{--}7 \times 10^{-2}$~kK$^{-1}$ (and
$\partial{\ln{S}}/\partial{\log{g}} \simeq -0.1$) in the $I$~band at
$\sim$30kK.  The magnitude of this dependence increases with
decreasing wavelength, indicating that for our purposes it is
preferable to work at $I$ rather than at $B$ or $V$.  (The effects of
extinction are also less at longer wavelengths.)

There is a significant ($\sim$10{\%}) dispersion in the $I$-band
surface brightness between models; with decreasing line blanketing the
surface brightness decreases (simply because more flux can escape at
shorter wavelengths).  At $\sim$30kK, the $I$-band surface of a
Howarth \& Lynas-Gray (SMC-metallicity) model is matched by that of an
{\sc atlas9} model $\sim$1kK cooler, and by a Smith \& Howarth model
$\sim$1kK hotter.

We adopt the Howarth \& Lynas-Gray model fluxes as the most
appropriate for our purposes.  While these LTE models are not suitable
for line-profile analyses, we expect them to give a reasonable
representation of the optical and near-IR continuum fluxes for
late-O/early-B main-sequence stars.

\begin{table*}
\caption{Photometric data.  The calculation of the absolute $I_{\rm
C}$ magnitudes at quadratures,
$M(I)$, is described in Section~\ref{metho_txt}.
The quoted $E(B-V)$ reddenings are calculated as
$E(B-I)/2.3$, as discussed in
Section~\ref{red_sec}.  $I_0$ is the dereddened $I$ magnitude at quadrature, 
and DM the true distance modulus.}
\label{absmag_tab}
\begin{tabular}{lccccccccc}
\hline
Star  & $M(I)$ & $B-I$ & $E(B-V)$ & $A(I)$ & $I_0$ & DM \\
\hline
5 038089 & $-3.87$ & $-0.37$ & 0.10 & 0.18 & 15.04 & 18.92 \\
5 202153 & $-5.05$ & $-0.36$ & 0.10 & 0.18 & 14.08 & 19.13 \\
5 316725 & $-4.36$ & $-0.40$ & 0.09 & 0.16 & 14.53 & 18.90 \\
6 077224 & $-4.98$ & $-0.28$ & 0.14 & 0.25 & 13.76 & 18.73 \\
6 158118 & $-4.00$ & $-0.24$ & 0.16 & 0.28 & 14.78 & 18.77 \\
6 215965 & $-4.86$ & $-0.41$ & 0.08 & 0.15 & 13.97 & 18.83 \\
7 243913 & $-4.52$ & $-0.25$ & 0.15 & 0.28 & 14.58 & 19.10 \\
9 175323 & $-5.31$ & $-0.41$ & 0.08 & 0.15 & 13.53 & 18.84 \\
11 30116 & $-4.01$ & $-0.37$ & 0.10 & 0.18 & 14.70 & 18.71 \\
11 57855 & $-3.03$ & $-0.45$ & 0.07 & 0.12 & 15.88 & 18.92 \\
\hline
\end{tabular}
\end{table*}

\subsection{Methodology and results}
\label{metho_txt}

The {\sc light2} code used for the light-curve analysis uses modified
black-body fluxes in its calculations, and is accordingly not suitable
for predicting stellar absolute magnitudes directly.  We therefore
took the best-fit parameters found by using {\sc light2}, and applied
them to a separate synthesis code, {\sc shade} (Howarth \& Wilson
1983).  This code incorporates essentially the same standard physics
as {\sc light2}, but allows more freedom in the choice of
model-atmosphere grids.\footnote{{\sc Shade} does not allow both stars
to be modelled simultaneously, which is why it could not be used for
the light-curve modelling.} Fluxes are interpolated linearly in
$\log{T_{\rm eff}}$ and $\log{g}$, in this case for grids sampled
every 2kK, 0.5~dex.  We converted from fluxes to magnitudes using
the zero-point calibrations given by Bessell et al.\ (1998; note that
the $f_\lambda$ and $f_\nu$ zero-points are transposed in their
Table~A2).

The procedure adopted was to match the mean radius of the primary, as
determined using {\sc light2}, and its mean temperature (assigned on
the basis of spectral type), and thereby to compute its $I$-band flux
at quadrature.  The radius of the secondary was then set at the {\sc
light2} value, and the temperature adjusted until the
primary:secondary quadrature flux ratio matched that determined by the
light-curve synthesis.  Since it is fundamentally the flux ratio, not
the relative temperatures, which is determined in the light-curve
synthesis, this is the most transparent and direct procedure for
determining the secondary temperatures and absolute fluxes.

The results on temperatures are included in
Table~\ref{light_curve_table} (where the errors on the secondary
temperatures propagate from the intensity ratios, and are therefore
{\em internal} errors), and the computed absolute magnitudes are given
in Table~\ref{absmag_tab}.

\begin{figure}
\includegraphics{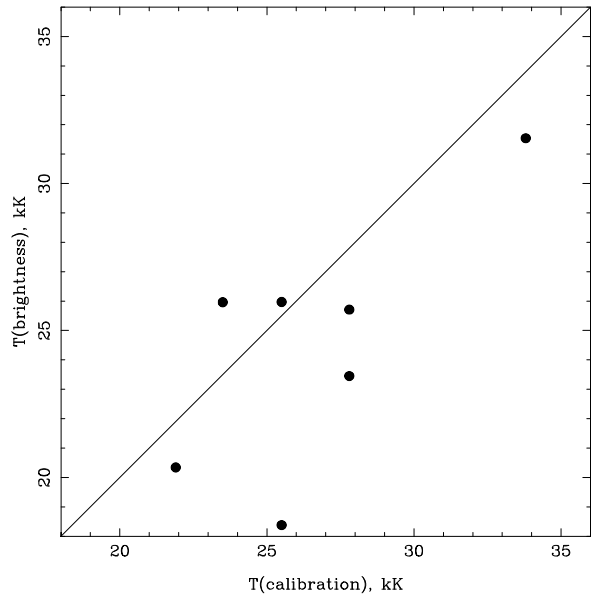}
\caption[]{Comparison of secondary-star effective temperatures inferred
from spectral types (`calibration' temperatures) and from surface
brightnesses.}
\label{tcomp_fig}
\end{figure}

\subsection{Temperatures revisited}

\label{tr_txt}

The secondary temperatures estimated from flux ratios may now be
compared with the temperatures implied by the spectral types.

The secondary temperatures required to match the observed
primary:secondary flux ratios appear systematically lower than the
temperatures implied by the spectral types, by $2.1 \pm 3.1$kK ($8
\pm 12$\%\, s.d.) on average.  That is, the temperature differences
between the primaries and secondaries as indicated by their spectral
types are, on average, smaller than the temperature differences
indicated by their surface-brightness ratios.  Obviously, a more
`compressed' temperature scale than the adopted one (i.e., a scale
in which the temperature difference between adjacent spectral types is
smaller) would exacerbate this problem.

The simplest (but not the only) interpretation is that the
secondaries' line spectra indicate spectral classifications $\sim$1
subtype earlier than do their $I$-band continuum fluxes.  A possible
explanation is that this is a consequence of the `reflection effect',
i.e., that the outer, line-forming regions of the secondary
atmospheres are heated by the incident radiation of the primary.

\section{Extinction}
\label{red_sec}

Interstellar reddening towards and within the SMC has been considered
by many authors, with a consensus that fairly small values of $E(B-V)$
generally apply.  Older studies are summarized by McNamara \& Feltz
(1980) and by Bessell (1991), who advocate foreground (and therefore
minimum) reddenings of $E(B-V) = 0.02$ and 0.05, respectively.  More
recent work indicates reddenings of $\sim$0.05--0.25.  (e.g., Hill,
Madore \& Freedman 1994; Massey et al.\ 1995; Dutra et al.\ 2001;
Zaritsky et al.\ 2002).  The {\sc ogle} data for our targets are in
good agreement with these results, as expected from the claimed
photometric accuracy of $\sim$0.01~mag (Udalski et al.\ 1998).

To investigate reddenings and extinctions to our targets, we assume
that the intrinsic colours of Galactic stars apply to our SMC targets
(the differences in these colours between Kurucz solar-metallicity and
SMC-metallicity models are negligible), and that the optical--near-IR
extinction curve is also similar (e.g., Feast \& Whitelock 1984;
Gordon \& Clayton 1998).

In principle, we want to use the longest available wavelength baseline
to determine reddenings -- i.e., $(B-I)$.   Writing
$x = E(V-I)/E(B-V)$ and $R_V = A(V)/E(B-V)$, in order to show explicitly
the dependence of extinction on these quantities, we then have
\begin{equation}
A(I) = \frac{E(B-I)}{1+x} (R_V-x).
\label{ai_eq}
\end{equation}
Unfortunately, although it is well established that
Galactic late-O/early-B near-main-sequence stars have $(B-V)_0 \simeq
-0.30$ (Fitzgerald 1970; Ducati et al.\ 2001), the intrinsic
$(V-I)$ colours are much more poorly determined.   We therefore
estimate $\overline{(V-I)}_0$ for our stars from the OGLE data 
by noting that
\[
(V-I)_0 = (V-I) - x\left((B-V) - (B-V)_0\right),
\]
finding $\overline{(V-I)}_0 = -0.30$ (i.e., $(B-I)_0 = -0.60$)
for $x = 1.30, (B-V)_0 = -0.30$.  This
compares favourably with values of $-0.28$ and $-0.33$ reported by
Wegner (1994) and by Ducati et al.\ (2001), after conversion from
$(V-I)_{\rm J}$ to $(V-I)_{\rm C}$ using the transformation given by Cousins
(1980).

The adopted $I$-band extinctions for our targets were
calculated by using eqtn.~\ref{ai_eq}, adopting $E(B-I) = (B-I) +
0.60$ together with $R_V = 3.1$, and $x = 1.30$ (Howarth 1983; Cardelli, Clayton \&
Mathis 1989; Mathis 1990).  Results are included in
Table~\ref{absmag_tab}.  Note that by adopting the $(B-I)$ reddenings
to determine the crucial quantity $A(I)$, we are not improving the
external {\em accuracy} of the results (which is established from the
$(B-V)$ colours); but that the internal {\em precision} is
considerably improved by using more than a threefold increase in
wavelength baseline.

The mean reddening towards our targets is $\overline{E}(B-V) = 0.11$,
with an s.d. of 0.03 (range 0.07--0.16).  A complementary indicator of
the consistency of the adopted reddenings can be obtained by comparing
two estimates of $E(B-V)$, viz.\ $(B-V)+0.30$ and $[(B-I)+0.60]/[1+x]$
for the sample.  The mean difference is zero (by construction), with
an s.d. of 0.021.

\section{The distance to the SMC}

The previous sections describe our determinations of the calculated
absolute $I$-magnitudes at quadratures (Section~\ref{metho_txt}) and
$I$-band extinctions (Section~\ref{red_sec}).  Combined with the
observed $I$-band quadrature magnitudes, these data yield the true
distance moduli directly.  The quadrature magnitudes come from the
normalizations of the light-curve models, and are included in
Table~\ref{basic_param_tab}; the distance moduli are given in
Table~\ref{absmag_tab}.  The average true distance modulus to our
targets, for the adopted temperature scale, is found to be $18.89 \pm
0.14$ (s.d.; error on the mean 0.04).

Although we can determine the distances to our targets with reasonable
precision, we need to consider their location within the SMC in order
to derive `the' distance.  The dynamics and structure of the SMC are
complex, probably as a result of interaction with the LMC (Murai \&
Fujimoto 1980; Gardiner \& Noguchi 1996), and that structure is,
arguably, still not completely clear.  Caldwell \& Coulson (1986),
Laney \& Stobie (1986), and Mathewson et al. (1988)
found line-of-sight depths of $\sim$20~kpc, while Welch et al.\ (1987)
concluded that the dispersion of the Cepheid
period--luminosity(--colour) relation corresponds to a depth of only
3.3kpc; Martin, Maurice \& Lequeux (1989) similarly concluded that
most of the young population lie within a band less than 10kpc deep,
as did Crowl et al.\ (2001).  Gardiner, Hatzidimitriou, Hawkins, and
collaborators have conducted a complementary series of studies of the
outer regions of the SMC (e.g., Hatzidimitriou, Cannon \& Hawkins
1993, and references therein).

Part of the discrepancy between different papers appears to arise
because of different interpretation of what `the' depth actually
means.  Our digest of published work is that the bulk of the young
populations at the centre of the SMC have a line of sight depth of
less than 10kpc; the main body of the bar is closer in the NE and more
distant in the SW, by $\sim$0.2 mag in distance modulus (Mathewson et
al.\ 1988; Gardiner \& Hawkins 1991).

Our target fields are less than 1$^\circ$ from the optical and
dynamical centres of the SMC.  The distance we find for our targets
should therefore be directly representative of the mean distance of
the SMC.  Moreover, the modest line-of-sight depth at this location
means that our sample should not be significantly biased towards
near-side objects (intrinsic luminosity, and variable transmission
through the 2dF fibre feeds, will be more important in determining
which targets have good S:N than will fore-and-aft location within the
Cloud).

Our distances are formally consistent with all targets lying at a
common distance. However, the two targets 5~202153 and 7~243913 are
distance outliers in the distribution, and hint at a real depth
effect, although their radial velocities, which might be expected to
reflect any distinct dynamical properties, are unexceptional
(Table~\ref{rv_tab}). Their discrepant distances cannot result
directly from the surface-brightness calibration, as these systems
have spectral types which fall in the main grouping of targets.

Finally, we note that our internal error estimates, discussed below,
are commensurate with the dispersion in distances to individual
systems, which provides the ultimate validation of our procedures.

\subsection{Error budget}

\subsubsection{Systematic errors}

Since reddenings are generally small, extinction is unlikely to be an
important source of systematic error; for example, the sensitivity of
true distance modulus to $R_V$, the ratio of selective to total
extinction, is $\partial{\mbox{\sc{dm}}}/\partial{R_V} = 0.1$ at $I$.
For SMC (internal) reddening of $E(B-V) = 0.1$, which is at the upper
end of the range found for our targets, even a systematic error in
$R_V$ as large as 0.5 (the extreme range of results given by Gordon \&
Clayton 1998) would change the true distance modulus by only 0.05.

Individual $E(B-V)$ reddening estimates are unlikely to be {\em
systematically} in error unless SMC stars of a given spectral type
have significantly different intrinsic colours to their Galactic
counterparts.  Model-atmosphere calculations, and observations,
provide no indication that this is the case.

The flux calibration of the magnitude scale (a source of systematic
error which affects all photometrically-based distance estimates) is
uncertain at the $\sim 1$\%\ level.

Potentially, our most important source of systematic error is the
adopted temperature--spectral-type scale (and the associated
relationship between surface brightness and effective temperature).
The spread in temperature scales is $\sim$1--2kK for late-O to early-B
types, which provides some guidance to the magnitude of systemetic
errors in this factor (although consistency may be a poor guide to
correctness).  For every 1kK change in $T_{\rm eff}$, there is a
$\sim$7\%\ change in surface brightness, or a $\sim$3\%\ change in
distance $d$ (0.06 mag in distance modulus).

Considering these factors, a reasonable estimate of the maximum
systematic error in the mean distance modulus from our (ultimately
large) sample is $\sim$0.1.  A hotter temperature scale would lead to
an increased distance estimate, as would (to a lesser extent) a
decrease in the SMC value of $R_V$.

\subsubsection{Random errors}
\label{random_sec}

By random errors we mean the statistical uncertainty on the estimated
distance to a single system.  This uncertainty results from errors on
the stellar radii; temperatures; and reddenings.

Since we essentially measure an angular diameter, errors on the radii
correlate directly with errors on distance.  These errors are
explicitly considered in the errors quoted in Table~\ref{light_curve_table};
a typical $\sim$4\%\ uncertainty in radius $R$ corresponds to an equal
uncertainty in $d$.

The spectral classifications of the primaries are estimated to be good
to $\sim$1 subtype; at O9-B0.5, this corresponds to an error of 
$\sim1\frac{1}{2}$kK in
$T_{\rm eff}$ which, as noted above, corresponds to a $\sim$5\%\ error
in distance.

The observational determination of quadrature magnitudes is not a
significant source of error (the light-curve normalizations are good
at the millimagnitude level).

Finally, an error of 0.02m in $E(B-V)$ corresponds to an error of 0.06
in distance modulus, or 3\%\ in $d$.

Adding these factors in quadrature, we conclude that the stochastic
error on the distance to a well-observed system is typically
$\sim$7\%\ (0.15 mag in distance modulus).

\begin{table}
\caption{Recent distance determinations to the SMC.}
\label{distances_tab}
\begin{tabular}{lll}
\hline
Method & DM & Author \\
\hline
\multicolumn{3}{c}{Period--luminosity-based methods} \\
IR Cepheid PL& $19.11\pm0.07$ & Visvanathan (1985) \\
Cepheid PL   & $18.97\pm0.07$ & Caldwell \& Coulson (1986) \\
Cepheid    PL& $18.93\pm0.05$ & Welch et al.\ (1987) \\
Cepheid    PL& $18.84\pm0.10$ & B\"ohm-vitense (1997) \\
Cepheid    PL& $19.11\pm0.11$ & Groenewegen (2000) \\
2$^{\rm nd}$ overtone Cep& $19.11\pm 0.11$ & Bono et al.\ (2001) \\
RR Lyr      &  $18.78 \pm 0.2$& Reid \& Strugnell (1986) \\
RR Lyr       & $18.66 \pm 0.16$ & Udalski (1998a) \\
\multicolumn{3}{c}{Red clump star methods} \\
Red clump & $18.56 \pm 0.06$ & Udalski et al.\ (1998a) \\
Red clump & $18.82 \pm 0.2$ & Cole (1998) \\
Red clump & $18.91 \pm 0.18$ & Twarog et al.\ (1999) \\
Red clump & $18.77 \pm 0.08$ & Popowski (2000) \\
\multicolumn{3}{c}{Other methods} \\
Eclipsing binary & $18.6\pm0.3$ & Bell et al.\ (1991) \\
Cepheid brightness & $18.9 \pm 0.2$ & Barnes et al.\ (1993) \\
Tip RGB & $18.99 \pm 0.08$ & Cioni et al.\ (2000) \\
CMD fitting & $18.88 \pm 0.08$ & Dolphin et al.\ (2001) \\
\hline
\end{tabular}
\end{table}

\section{Discussion}

The distance to the SMC has been measured using a variety of methods,
although the most frequently adopted techniques are the Cepheid
period--luminosity (PL) relation and, more recently, the red-clump
method (see Table~\ref{distances_tab}). The Cepheid period-luminosity
relation has the advantage that the observations themselves are
relatively straightforward, but the method requires a zero-point
calibration (there are no reliable {\it Hipparcos} parallaxes of
Cepheids; e.g., Madore \& Freedman 1998), and the PL relation may have
a metallicity dependence (e.g., Sasselov et al.\ 1997; Caputo et al.\ 
2000).  Despite these uncertainties, and although it does not provide
a {\em primary} distance indicator, the Cepheid method remains a
crucially important rung in the cosmic distance ladder, and is the
foundation upon which the Hubble Key Project builds its measurement of
$H_0$ (Freedman et al.\ 2001). Recent determinations of the distance
modulus to the SMC using the PL relationship range from 18.84
(B\"ohm-Vitense 1997) to 19.11 (Visvanathan 1985; Groenewegen 2000;
Bono, Caputo \& Marconi 2001), and, broadly speaking, establish the
`long' distance scale.

The red-clump (RC) stars are metal-rich analogues of horizontal-branch
stars, whose luminosity is thought to be independent of their age and
chemical composition (to first order), so that they may be used as
standard candles. One of the advantages of RC method is that it is
developed from reliable {\it Hipparcos} parallaxes.  The first
application of the RC method to the SMC yielded a surprisingly short
DM of 18.6 (Udalski et al.\ 1998a).  Later studies by Cole (1998) and
Twarog et al. (1999) invoked age and metallicity dependencies of the
RC stars, yielding slightly longer distances
(Table~\ref{distances_tab}), but still somewhat shorter than the
Cepheid scale.  Although Udalski (1998b) proposed that the $I$-band
metallicity dependence is weak, and that the RC luminosity is
independent of age for stars 2--10\,Gyr old, modeling of RC
populations (Girardi \& Salaris 2001) suggests that the situation may
be complex, with a full calibration of the RC distance scale requiring
knowledge of the star-formation history and the age--metallicity
relation.

\begin{figure}
\includegraphics{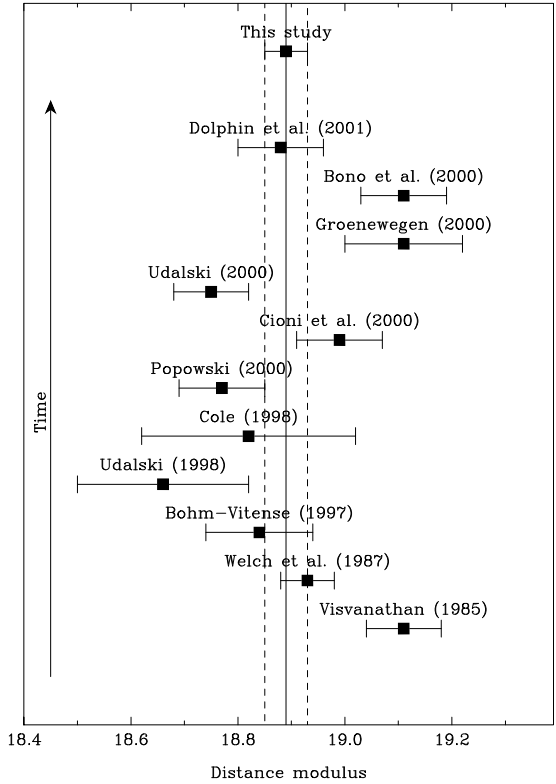}
\caption{Distance moduli to the SMC calculated using a variety of
  techniques (see Table~\protect\ref{distances_tab}). The DM are
  plotted as filled squares with $1\sigma$-error bars. The distance
  modulus derived from this study is shown for comparison with the
  dotted lines delineating the $1\sigma$ random error limit.}
\label{dist_fig}
\end{figure}

The distance modulus derived here ($18.89\pm0.04$ mag) bisects the
`short' (red clump) and `long' (Cepheid) characterisitic distances,
and is in good agreement with, but more precise than, the Dolphin et
al. (2001) measurement from CMD fitting (see Fig.~\ref{dist_fig} for a
graphical comparison of recent distance estimates).  We are continuing
with the analysis of the remainder of our sample, although further
spectroscopic observations will be necessary for complete quadrature
phase coverage of all our objects. The final result of this programme
will be a substantial database of stellar parameters with which to
test stellar evolution models (both single and binary). We will also
be able to refine our distance estimate, and to examine further the
three-dimensional structure of the central regions of the SMC.

\section*{Acknowledgements}
This paper is based on data obtained by using the Anglo-Australian
Telescope. Terry Bridges is warmly thanked for his expert help with
the 2dF astrometry, for conducting the observations, and for his data
reduction advice. We are grateful to the OGLE team for making their
excellent photometric database publically available, and Graham~Hill
for the use of his {\sc light2} code.

\section*{Appendix A}

 \begin{table*}
 \begin{center}
 \caption{Radial-velocity measurements. The first column lists
   the heliocentric julian date of mid-exposure, while $v_p$ and $v_s$
   denote the primary and secondary radial velocities respectively.
   The errors quoted were determined from the fit covariance matrix.}
 {\scriptsize
\begin{tabular}{crrrrrrrrrr}
\hline
 HJD       &   \multicolumn{1}{c}{$v_p$}        &     \multicolumn{1}{c}{$v_s$}      &   \multicolumn{1}{c}{$v_p$}        &     \multicolumn{1}{c}{$v_s$}     &   \multicolumn{1}{c}{$v_p$}         &     \multicolumn{1}{c}{$v_s$}      &  \multicolumn{1}{c}{$v_p$}        &     \multicolumn{1}{c}{$v_s$}      &   \multicolumn{1}{c}{$v_p$}        &     \multicolumn{1}{c}{$v_s$}       \\
 $-240000$ &    (\kms)      &     (\kms)     &    (\kms)      &     (\kms)    &    (\kms)       &     (\kms)     &    (\kms)      &     (\kms)     &    (\kms)      &     (\kms)      \\
\hline
           & \multicolumn{2}{c}{\bf 5 038089}    & \multicolumn{2}{c}{\bf 5 202153}   & \multicolumn{2}{c}{\bf 5 316725}     & \multicolumn{2}{c}{\bf 6 077224}    & \multicolumn{2}{c}{\bf 6 158118}     \\
52158.2834 &$   +412 \pm \ph{0}6$&$   \ph{00}-3\pm 11 $&$   +282 \pm 14$&$     \ph{00+}0\pm 10 $&$    \ph{0}+24 \pm 12$&$   +436\pm 12 $&                          &                &$   +309 \pm 16$&$    \ph{0}-62\pm 19 $ \\
52158.9872 &                &                &                &                &                &                &                $    \ph{0}+33 \pm 10$ &$   +367\pm \ph{0}9 $&                 &                 \\
52159.0346 &                &                &                &                &                &                &$    \ph{0}+30 \pm 11$&$   +380\pm 10 $&                &                 \\
52159.0641 &                &                &                &                &                &                &$    \ph{0}+22 \pm 10$&$   +389\pm 10 $&                &                 \\
52159.1007 &                &                &                &                &                &                &                &                &                &                 \\
52159.1587 &                &                &                &                &                &                &$    \ph{0}+22 \pm 11$&$   +373\pm 10 $&$    \ph{0}+75 \pm 13$&$   +402\pm 21 $ \\
52159.1886 &                &                &$   +119 \pm 14$&$   +259\pm 11 $&                &                &$    \ph{0}+29 \pm 11$&$   +375\pm \ph{0}9 $&$    \ph{0}+64 \pm 20$&$   +418\pm 23 $ \\
52159.2430 &                &                &$    \ph{0}+80 \pm 14$&$   +295\pm \ph{0}3 $&                &                &$    \ph{0}+22 \pm 11$&$   +369\pm 10 $&$    \ph{0}+30 \pm 18$&$   +424\pm 10 $ \\
52160.9686 &                &                &                &                &                &                &$   +335 \pm 10$&$    \ph{0}-23\pm \ph{0}9 $&                &                 \\
52161.0089 &                &                &                &                &$    \ph{0}+17 \pm 13$&$   +457\pm 10 $&$   +358 \pm 11$&$    \ph{0}-16\pm \ph{0}9 $&                &                 \\
52161.0992 &$   +414 \pm 13$&$   -002\pm 12 $&                &                &$    \ph{0}+18 \pm 14$&$   +441\pm 12 $&$   +340 \pm 11$&$     \ph{00}-7\pm 10 $&                &                 \\
52161.1495 &                &                &                &                &$    \ph{0}+28 \pm 15$&$   +428\pm 13 $&$   +332 \pm 11$&$    \ph{0}-15\pm 10 $&                &                 \\
52161.1798 &                &                &                &                &$    \ph{0}+40 \pm 16$&$   +422\pm 12 $&$   +333 \pm 11$&$     \ph{00}-9\pm 10 $&                &                 \\
52161.2097 &                &                &                &                &                &                &$   +335 \pm 11$&$    \ph{0}-12\pm 10 $&                &                 \\
52161.2380 &                &                &                &                &                &                &$   +333 \pm 11$&$    \ph{0}-15\pm 10 $&                &                 \\
52161.2667 &                &                &                &                &                &                &$   +318 \pm 11$&$    \ph{0}-15\pm 10 $&                &                 \\
52161.2912 &                &                &                &                &                &                &$   +339 \pm 11$&$     \ph{00}-5\pm 10 $&                &                 \\
52161.9853 &                &                &$   +321 \pm \ph{0}7$&$    \ph{0}-60\pm 10 $&                &                &                &                &$    \ph{0}+66 \pm 16$&$   +472\pm 22 $ \\
52162.0273 &$   \ph{0}-71 \pm 14$&$   +421\pm \ph{0}7 $&$   +312 \pm \ph{0}5$&$    \ph{0}-65\pm \ph{0}9 $&$   +307 \pm 16$&$    \ph{0}-71\pm 13 $&                &                &$    \ph{0}+61 \pm 17$&$   +478\pm 21 $ \\
52162.0714 &$   \ph{0}-74 \pm 14$&$   +415\pm \ph{0}7 $&$   +298 \pm 14$&$    \ph{0}-68\pm 10 $&$   +315 \pm 12$&$    \ph{0}-88\pm 13 $&                &                &$    \ph{0}+86 \pm 17$&$   +479\pm 21 $ \\
52162.1128 &$   \ph{0}-66 \pm 13$&$   +416\pm \ph{0}6 $&$   +309 \pm 12$&$    \ph{0}-86\pm 11 $&$   +308 \pm 16$&$    \ph{0}-99\pm 13 $&                &                &$    \ph{0}+58 \pm 18$&$   +421\pm 15 $ \\
52162.1506 &$   \ph{0}-69 \pm 15$&$   +411\pm 11 $&$   +317 \pm \ph{0}3$&$    \ph{0}-84\pm 11 $&$   +324 \pm 10$&$   -114\pm 12 $&                &                &$    \ph{0}+37 \pm 18$&$   +430\pm 11 $ \\
52162.1795 &$   \ph{0}-67 \pm 15$&$   +407\pm 13 $&$   +298 \pm 14$&$    \ph{0}-96\pm 11 $&$   +316 \pm 11$&$   -121\pm 12 $&                &                &$    \ph{0}+71 \pm 12$&$   +430\pm 13 $ \\
52162.2091 &$   \ph{0}-61 \pm 15$&$   +402\pm 13 $&$   +314 \pm \ph{0}4$&$    \ph{0}-79\pm 10 $&$   +323 \pm \ph{0}9$&$   -124\pm 12 $&                &                &$    \ph{0}+76 \pm 14$&$   +427\pm \ph{0}9 $ \\
52162.2375 &$   \ph{0}-53 \pm 15$&$   +395\pm 13 $&$   +317 \pm \ph{0}3$&$    \ph{0}-92\pm 11 $&$   +318 \pm \ph{0}8$&$   -120\pm 12 $&                &                &$    \ph{0}+72 \pm 12$&$   +414\pm 21 $ \\
52162.2674 &$   \ph{0}-48 \pm 14$&$   +388\pm \ph{0}9 $&$   +312 \pm \ph{0}6$&$    \ph{0}-84\pm 11 $&$   +318 \pm \ph{0}8$&$   -137\pm 12 $&                &                &                &                 \\
52162.9674 &                &                &$   +250 \pm 14$&$    \ph{0}+17\pm 10 $&                &                &$    \ph{0}+26 \pm 11$&$   +369\pm \ph{0}9 $&                &                 \\
52163.0101 &                &                &$   +270 \pm 13$&$    \ph{0}+33\pm 10 $&                &                &$     \ph{00}-8 \pm 10$&$   +381\pm \ph{0}9 $&                &                 \\
52163.0531 &$   +415 \pm 12$&$   \ph{00}-6\pm 10 $&$   +281 \pm 14$&$    \ph{0}+61\pm 10 $&                &                &$    \ph{0}+28 \pm 11$&$   +382\pm 10 $&$   +311 \pm 17$&$    \ph{0}-49\pm 22 $ \\
52163.0820 &$   +426 \pm 12$&$   \ph{00}+9\pm 12 $&$   +248 \pm 14$&$    \ph{0}+68\pm \ph{0}8 $&                &                &                &                &$   +310 \pm 17$&$    \ph{0}-54\pm 22 $ \\
52163.1104 &$   +428 \pm 14$&$   \ph{0}-16\pm 12 $&$   +255 \pm 14$&$    \ph{0}+73\pm 10 $&                &                &$    \ph{0}+36 \pm 11$&$   +377\pm \ph{0}9 $&                &                 \\
52163.1405 &                &                &                &                &                &                &$    \ph{0}+23 \pm 11$&$   +364\pm 10 $&                &                 \\
 \\
           & \multicolumn{2}{c}{\bf 6 215965}    & \multicolumn{2}{c}{\bf 7 243913}    & \multicolumn{2}{c}{\bf 9 175323}    & \multicolumn{2}{c}{\bf 11 30116}    & \multicolumn{2}{c}{\bf 11 57855}    \\
52158.2834 &                &                &$   +356 \pm 17$&$    \ph{0}-67\pm 20 $&                &                &                &                &                                 \\ 
52158.9872 &$    \ph{0}-16 \pm 12$&$   +358\pm 11 $&                &                &$   +383 \pm 12$&$    \ph{0}-41\pm 11 $&$   +279 \pm 14$&$    \ph{0}+61\pm 15 $&$   +356 \pm 14$&$    \ph{0}-97\pm 15 $\\
52159.0346 &$    \ph{0}-13 \pm 11$&$   +375\pm \ph{0}9 $&                &                &                &                &$   +295 \pm 11$&$    \ph{0}+29\pm 11 $&                &                \\
52159.0641 &$    \ph{0}-12 \pm 11$&$   +384\pm 10 $&                &                &                &                &                &                &$   +352 \pm 15$&$   -109\pm 18 $\\
52159.1007 &$    \ph{0}-29 \pm 12$&$   +361\pm 11 $&                &                &                &                &                &                &                &                \\
52159.1587 &$    \ph{0}-25 \pm 11$&$   +381\pm 11 $&                &                &                &                &$   +313 \pm 15$&$    \ph{0}-18\pm 16 $&$   +300 \pm \ph{0}7$&$    \ph{0}-68\pm 19 $\\
52159.1886 &$    \ph{0}-44 \pm 12$&$   +375\pm 10 $&                &                &                &                &                &                &                &                \\
52159.2430 &$    \ph{0}-39 \pm 12$&$   +373\pm 10 $&$    \ph{0}+60 \pm 17$&$   +425\pm 20 $&                &                &$   +326 \pm 15$&$    \ph{0}-32\pm 16 $&$   +259 \pm 14$&$    \ph{0}+22\pm 18 $\\
52160.9686 &$   +365 \pm \ph{0}9$&$    \ph{0}-18\pm 10 $&$   +343 \pm 17$&$    \ph{0}-56\pm 19 $&$   +399 \pm 13$&$    \ph{0}-57\pm 13 $&$    \ph{0}+65 \pm 13$&$   +475\pm 16 $&$    \ph{0}-23 \pm 15$&$   +459\pm 20 $\\
52161.0089 &$   +321 \pm 11$&$    \ph{0}-34\pm \ph{0}9 $&                &                &                &                &                &                &                &                \\
52161.0992 &$   +360 \pm \ph{0}7$&$    \ph{0}-37\pm \ph{0}9 $&                &                &                &                &                &                &                &                \\
52161.1495 &$   +367 \pm 11$&$    \ph{0}-47\pm 11 $&                &                &$   +392 \pm 13$&$    \ph{0}-59\pm 13 $&$    \ph{0}+70 \pm 15$&$   +432\pm 16 $&$    \ph{0}+43 \pm 14$&$   +356\pm 18 $\\
52161.1798 &$   +361 \pm \ph{0}7$&$    \ph{0}-50\pm 10 $&                &                &                &                &                &                &                &                \\
52161.2097 &$   +352 \pm 12$&$    \ph{0}-44\pm 11 $&                &                &$   +379 \pm 13$&$    \ph{0}-31\pm 13 $&$    \ph{0}+80 \pm 15$&$   +414\pm 13 $&                &                \\
52161.2380 &$   +355 \pm 10$&$    \ph{0}-54\pm 11 $&                &                &                &                &                &                &                &                \\
52161.2667 &$   +366 \pm \ph{0}9$&$    \ph{0}-55\pm 10 $&                &                &                &                &                &                &                &                \\
52161.2912 &$   +369 \pm 11$&$    \ph{0}-37\pm 10 $&                &                &                &                &                &                &                &                \\
52161.9853 &                &                &$    \ph{0}+54 \pm 20$&$   +466\pm 20 $&$    \ph{0}+50 \pm 12$&$   +454\pm \ph{0}8 $&$   +287 \pm \ph{0}9$&$    \ph{0}+44\pm 16 $&                &                \\
52162.0273 &                &                &$    \ph{0}+48 \pm 13$&$   +478\pm 11 $&                &                &                &                &                &                \\
52162.0714 &                &                &$    \ph{0}+49 \pm 15$&$   +479\pm 11 $&$    \ph{0}+33 \pm 11$&$   +478\pm 13 $&$   +308 \pm 15$&$     \ph{00}+2\pm 16 $&                &                \\
52162.1128 &                &                &$    \ph{0}+34 \pm 14$&$   +480\pm 11 $&                &                &                &                &                &                \\
52162.1506 &                &                &$    \ph{0}+45 \pm 20$&$   +482\pm 20 $&$    \ph{0}+17 \pm 13$&$   +484\pm 13 $&$   +316 \pm 15$&$    \ph{0}-19\pm 16 $&$     \ph{00}+7 \pm 14$&$   +414\pm \ph{0}6 $\\
52162.1795 &                &                &$    \ph{0}+25 \pm 16$&$   +473\pm 10 $&                &                &                &                &                &                \\
52162.2091 &                &                &$    \ph{0}+41 \pm 12$&$   +468\pm 19 $&$    \ph{0}+20 \pm 13$&$   +485\pm 13 $&$   +320 \pm 15$&$    \ph{0}-37\pm 15 $&$    \ph{0}-11 \pm 11$&$   +447\pm 19 $\\
52162.2375 &                &                &$    \ph{0}+36 \pm 18$&$   +464\pm 23 $&                &                &                &                &                &                \\
52162.2674 &                &                &$    \ph{0}+49 \pm 14$&$   +466\pm 17 $&$    \ph{0}+25 \pm 14$&$   +473\pm 14 $&$   +328 \pm 15$&$    \ph{0}-51\pm 16 $&$    \ph{0}-17 \pm 15$&$   +448\pm 20 $\\
52162.9674 &$    \ph{0}-23 \pm 11$&$   +348\pm 11 $&                &                &                &                &$   +292 \pm \ph{0}9$&$    \ph{0}+96\pm 16 $&$   +366 \pm 14$&$   -112\pm 19 $\\
52163.0101 &$    \ph{0}-22 \pm 11$&$   +368\pm 10 $&                &                &                &                &                &                &                &                \\
52163.0531 &$    \ph{0}-17 \pm 11$&$   +372\pm 11 $&                &                &$   +380 \pm 11$&$     \ph{00}-1\pm 12 $&                &                &$   +313 \pm 14$&$    \ph{0}-55\pm 10 $\\
52163.0820 &$    \ph{0}-19 \pm 11$&$   +372\pm 10 $&                &                &                &                &                &                &                &                \\
52163.1104 &$    \ph{0}-30 \pm 12$&$   +372\pm 11 $&                &                &$   +395 \pm 12$&$    \ph{0}-22\pm 12 $&                &                &$   +280 \pm 14$&$     \ph{00}-5\pm 19 $\\
52163.1405 &$    \ph{0}-40 \pm 12$&$   +379\pm \ph{0}9 $&                &                &                &                &                &                &                &                \\
\hline
\end{tabular}
}
\end{center}
\end{table*}

\end{document}